\documentclass[10pt,a4paper]{article}
\usepackage[utf8]{inputenc}
\usepackage[english]{babel}
\usepackage{amsmath}
\usepackage{amsfonts}
\usepackage{amssymb}
\usepackage[left=2.5cm,right=2.5cm,top=3cm,bottom=2.5cm]{geometry}
\usepackage{graphicx}
\usepackage{xcolor}
\usepackage{placeins}
\usepackage{multirow}
\usepackage{rotating}
\usepackage[normalem]{ulem}
\usepackage{algorithm}
\usepackage{algpseudocode}
\usepackage{titlesec}
\title{A discontinuous Galerkin approach for simulating graphene-based electron devices via the Boltzmann transport equation}
\author{Giovanni Nastasi$^1$ \qquad Vittorio Romano$^2$}
\date{$^1$University of Enna ``Kore'', Department of Engineering and Architecture\\
Plesso E - Cittadella Universitaria, 94100 - Enna, Italy\\
\texttt{giovanni.nastasi@unikore.it}\\
\medskip
$^2$University of Catania, Department of Mathematics and Computer Science\\
Viale A. Doria, 6, 95125 - Catania, Italy\\
\texttt{vittorio.romano@unict.it}}

\newcommand{\bx}{\mathbf{x}}
\newcommand{\bk}{\mathbf{k}}
\newcommand{\te}{t}
\newcommand{\R}{\mathbb{R}}
\newcommand{\eps}{\varepsilon}
\newcommand{\dm}{\displaystyle}
\newcommand{\norm}[1]{\left\lVert #1 \right\rVert}

\newcommand{\bibnam}[2]{#1 #2,}
\newcommand{\bibart}[4]{#1, #2, #3 (#4).}
\newcommand{\bibbook}[3]{#1, #2 (#3).}

\begin{document}
\maketitle

\begin{abstract}
Electron devices based on graphene have lately received a considerable interest; in fact, they could represent the ultimate miniaturization,  since the active area is only one atom thick. However, the gapless dispersion relation of graphene at the Dirac points limits the possibility of using pristine graphene instead of traditional semiconductors in 
Field Effect Transistors (FET). For such a reason very accurate simulations are needed. 

In Nastasi and Romano (IEEE Trans. Electron Devices 68, 2021) a graphene field effect transistor (GFET) has been proposed and simulated adopting a drift-diffusion model. Here, electron devices whose active area is made of monolayer graphene are simulated taking as mathematical model the complete semiclassical Boltzmann transport equation (BTE) retaining the full complexity of the collision operator which includes all the electron-phonon scatterings along with the Pauli exclusion principle. The transport equation is coupled with the Poisson equation for the electric field. 

The Boltzmann equation is solved by means of a discontinuous Galerkin (DG) approach 
with linear elements in space  and piecewise constant approximation in the wave-vector. The correct physical range for the distribution function is preserved with a maximum-principle-preserving scheme. Finite differences for
the Poisson equation, and a TVD Runge-Kutta scheme for time discretization are adopted.
The method reveals very robust and possesses a good degree of accuracy, making it particularly well suited for capturing the complex charge transport dynamics inherent to graphene-based devices.
The kinetic approach reveals new features which are caught by the standard drift-diffusion equations; in particular, the presence of a boundary layer at the contacts.
\end{abstract}

\bigskip


\medskip


\section{Introduction}
Semiconductor materials are foundational to modern electronic devices, employed in all the technological applications. As the demand for more powerful yet energy-efficient technology grows, the field of semiconductor research requires to enhance device performance while simultaneously reducing energy consumption and mitigating environmental impact. The pursuit of greater miniaturization remains a central pillar of semiconductor innovation. Furthermore, the search for next-generation materials has led to the emergence of low-dimensional materials as a transformative area of study. 

Graphene exists in a pristine form as a large area single atomic layer. It exhibits extraordinary electronic properties that could surpass the capabilities of traditional silicon-based components. The unique characteristics of this material, including high electron mobility and tunable bandgaps as in nanoribbons, position it as potential replacements for conventional semiconductors, paving the way for a new era of highly efficient and high-performance electronic devices \cite{CaNe}. 

By large-area graphene, we refer to samples whose lateral dimensions typically range from a few micrometers up to several millimeters, depending on the fabrication technique. For instance, mechanically exfoliated graphene flakes are usually of the order of $1$--$100~\mu\text{m}$, while chemical vapor deposition (CVD) methods can produce continuous graphene sheets with lateral sizes up to the millimeter or even wafer scale.
In the present work, the simulated device dimensions are chosen within the micrometer range, which is consistent with typical GFET channel lengths reported in the literature \cite{GeNo}.

A class of devices of this kind is represented by graphene field effect transistors (GFETs), see \cite{Schwierz} for a comprehensive review. Such devices are commonly investigated using reduced one-dimensional models, often employing averaging procedures to simplify the description of the active region \cite{Jimenez,Upadhyay}. Charge transport in such models is typically treated within the drift-diffusion framework. Full coupled drift-diffusion-Poisson simulations of GFETs have been proposed in \cite{NaRo_CNSNS,Jourdana}. As an alternative, hydrodynamical models have been explored to capture additional transport phenomena \cite{Bar,Ba,LuRo,LuRo2,CaRoVi}, and in several studies thermal effects have also been incorporated  \cite{MaRo}.

A widely used deterministic approach able to numerically solve many different kinds of equations is the Discontinuous Galerkin (DG) method \cite{BOOK:Hesthaven}. In \cite{CoShu} the DG method has been adapted to hyperbolic problems by using a Runge-Kutta integration in time. In this work,  extending previous studies performed in the spatially homogeneous case \cite{MajNaRo}, we directly solve the semiclassical Boltzmann equation for electrons in graphene coupled to the Poisson equation for the electrostatic potential by using a DG method \cite{Maj,Majo}. A first attempt has been presented in \cite{NaRo_RicMat}, where in the case of an assigned potential constant elements have been adopted. Here,  
linear elements in the spatial coordinate and constant  in the wave-vector space
are used. The correct physical range for the distribution function is preserved with a maximum-principle-preserving scheme. The Poisson equation is discretized with finite differences and a TVD Runge-Kutta scheme is adopted for time discretization.  A major issue is the definition of numerical fluxes at the interface between two different elements. This problem is solved  with an appropriate generalization of the numerical flux in the piece-wise constant case. The readers interested in finite-difference-based schemes are referred to \cite{LMS}.

Simulation of suspended and contacted monolayer graphene and a GFET show a good robustness and accuracy of the proposed scheme.

The plan of the paper is as follows. In Sec.  \ref{sec:Boltzmann} the semiclassical Boltzmann transport equation for charge transport in graphene is introduced including all the relevant electron-phonon scatterings and scattering with impurities. In Sec. \ref{sec:setting} the simulated devices are described:  specifically, a suspended contacted monolayer graphene, and a GFET similar to those in \cite{NaRo_IEEE}. Sec. \ref{sec:numerics} is devoted to the numerical scheme while in the last section  the numerical results are shown and discussed.

\section{The semiclassical Boltzmann model} \label{sec:Boltzmann}
The semiclassical Boltzmann equation is considered as the most accurate model for charge transport in electron devices, except peculiar situations where quantum effects are relevant, e.g. in resonant tunneling diodes. In the case of two dimensional materials, and specifically in graphene,  they read in the bipolar case as
\begin{equation}\label{EQ:Boltzmann}
\frac{\partial f_s}{\partial t} + \mathbf{v}_s \cdot \nabla_{\bx} f_s -\frac{e}{\hbar}\mathbf{E}\cdot\nabla_{\bk} f_s = Q(f_s,f_{-s}), \qquad s=\pm,
\end{equation}
where $f_s=f_s(t,\bx,\bk)$ is the electron distribution function in the conduction band ($s=+$) or in the valence band ($s=-$) at time $t>0$, position $\bx\in D\subset\mathbb{R}^2$ and wave-vector $\bk\in\mathcal{B}\subset\mathbb{R}^2$; $-s$ means sign reversal; $D$ is the domain representing the active area of the device and $\mathcal{B}$ is the first Brillouin zone. Indeed, $\mathcal{B}$ is usually  extended to $\mathbb{R}^2$ by homogenization \cite{BOOK:Jacoboni}. In \eqref{EQ:Boltzmann} $e$ represents the elementary charge and $\hbar$ is the reduced Planck constant. The group velocity $\mathbf{v}_s $ is given by
\begin{equation}
\mathbf{v}_s = \frac{1}{\hbar}\nabla_{\bk}\varepsilon_s (\bk),
\end{equation}
where $\varepsilon_s (\bk) $ is the dispersion relation of electrons. For pristine graphene one has  
$\varepsilon_s=s\hbar v_F\vert \bk \vert$, $v_F$ being the Fermi velocity which is a constant quantity.

The electric field $\mathbf{E}=\mathbf{E}(t,\bx) = (E_x(t,\bx), E_y(t,\bx)) $ is related to the electrostatic potential $\phi(t,\bx)$ by
$$
\mathbf{E}=-\nabla_{\bx}\phi,
$$
with $\phi$ obtained by coupling eq.s \eqref{EQ:Boltzmann} by the Poisson equation
\begin{equation}\label{EQ:Poisson}
\nabla\cdot(\epsilon\nabla_{\bx} \phi) = h,
\end{equation}
where $\epsilon$ is the dielectric constant and $h$ the carrier density. Indeed, they depend on the specific geometry of the device and will be discussed in Sec. \ref{SEC:Phys_sett}. 

The collision term $Q(f_s,f_{-s})$ includes the scatterings between electrons and phonons in the graphene, and  the scattering between electrons and impurities and phonons in the  oxide substrate,
if present,
\begin{equation*}
Q( f_{s}, f_{-s} ) = Q^{(el-ph)}( f_{s}, f_{-s} ) + Q^{(el-sub)}( f_{s}, f_{-s} ) .
\end{equation*}

The term $Q^{(el-ph)}$ consists of several contributions: the interaction  of electrons with acoustic, optical and $K$ phonons of graphene. Acoustic phonon scattering (ac) occurs within the same valley and band (intra-valley and intra-band). Optical phonon scattering also takes place within the same valley (intra-valley) and can involve either longitudinal optical (LO) or transverse optical (TO) phonons. It may be intra-band, where the electron remains in the same energy band, or inter-band, where the electron transitions to a different band. Scattering with optical phonons of type $K$ results in inter-valley transitions, moving electrons between adjacent valleys. If there is an oxide, remote optical phonon scattering, which is also intra-valley, can involve LO-sub and TO-sub phonons and may be intra-band or inter-band. Similarly, remote impurity scattering (imp) is limited to intra-valley and intra-band processes. In all cases, phonons are assumed to be in thermal equilibrium.
Therefore, the general form of $Q^{(el-ph)}$ is
\begin{align*}
Q^{(el-ph)}( f_{s}, f_{-s} ) = &
\sum_{s'} \left[ \int_{\R^2} S_{s', s}^{(el-ph)}(\bk', \bk) \, f_{s'}(\te,\bx,\bk') \left( 1 - f_{s}(\te,\bx,\bk) \right) d \bk'  \right. \\
& \left. - \int _{\R^2}S_{s, s'}^{(el-ph)}(\bk, \bk') \, f_{s}(\te,\bx,\bk) \left( 1 - f_{s'}(\te,\bx,\bk') \right) d \bk' \right],
\end{align*}
where the total transition rate is given by the sum of the contributions of the several types of scatterings
\begin{equation}\label{Scatt}
\begin{aligned}
S_{s', s}^{(el-ph)}(\bk', \bk) =  \sum_{\nu} \left| G^{(\nu)}_{s', s}(\bk', \bk) \right|^{2} & \left[ \left( n^{(\nu)}_{\mathbf{q}} + 1 \right) \delta \left( \eps_{s}(\bk) - \eps_{s'}(\bk') + \hbar \, \omega^{(\nu)}_{\mathbf{q}} \right) \right. \\
& \left.  + n^{(\nu)}_{\mathbf{q}} \,
\delta \left( \eps_{s}(\bk) - \eps_{s'}(\bk') - \hbar \, 
\omega^{(\nu)}_{\mathbf{q}}
\right) \right] . 
\end{aligned}
\end{equation}
The index $\nu$ labels the $\nu$th phonon mode, $\left| G^{(\nu)}_{s', s}(\bk', \bk) \right|$ is the 
matrix element, which describes the scattering mechanism, due to phonons of type $\nu$, between 
electrons belonging to the band $s'$ and electrons belonging to the band $s$.
The symbol $\delta$ denotes the Dirac distribution, $\omega^{(\nu)}_{\mathbf{q}}$ is the
the $\nu$th  phonon  frequency, $n^{(\nu)}_{\mathbf{q}}$ is the Bose-Einstein distribution for 
the phonon of type $\nu$
$$
n^{(\nu)}_{\mathbf{q}} = \dfrac{1}{e^{\hbar \, \omega^{(\nu)}_{\mathbf{q}} /k_B T} - 1},
$$
$k_B$ being the Boltzmann constant and $T$ being the  graphene lattice temperature which, in this article, is assumed constant.
When, for a phonon $\nu_{*}$, $\hbar \, \omega^{(\nu_{*})}_{\mathbf{q}} \ll k_B T$, then the 
scattering with the phonon $\nu_{*}$ can be assumed elastic. 
In this case, we eliminate in Eq.~\eqref{Scatt} the term 
$\hbar \, \omega^{(\nu_{*})}_{\mathbf{q}}$ inside the delta distribution and we use the 
approximation $n^{(\nu_{*})}_{\mathbf{q}} + 1 \approx n^{(\nu_{*})}_{\mathbf{q}}$.

Now we write explicitly the transition  rates used in our simulations. For acoustic phonons, usually one considers the elastic approximation for which the following relation holds 
\begin{equation}\label{transport_acoustic}
2 \, n^{(ac)}_{\mathbf{q}}
\left| G^{(ac)}(\bk', \bk) \right|^{2} = \dfrac{1}{(2 \, \pi)^{2}} \, 
\dfrac{\pi \, D_{ac}^{2} \, k_{B} \, T}{2 \hbar \, \sigma_m \, v_{p}^{2}}
\left( 1 + \cos \vartheta_{\bk, \bk'} \right) ,
\end{equation}
where $D_{ac}$ is the acoustic phonon coupling constant, $v_{p}$ is the sound speed in 
graphene, 
$\sigma_m$ the graphene areal density, 
and $\vartheta_{\bk, \bk'}$ is the convex angle between $\bk$ and ${\bk'}$.   
\\
There are three relevant optical phonon scatterings: the longitudinal optical (LO), the 
transversal optical (TO) and the ${K}$  phonons.
The matrix elements read
\begin{align}
&
\left| G^{(LO)}(\bk', \bk) \right|^{2} + \left| G^{(TO)}(\bk', \bk) \right|^{2} =
\dfrac{2}{(2 \, \pi)^{2}} \, \dfrac{\pi \, D_{O}^{2}}{\sigma_m \, \omega_{O}},
\\
&
\left| G^{(K)}(\bk', \bk) \right|^{2} = 
\dfrac{1}{(2 \, \pi)^{2}} \, \dfrac{2 \pi \, D_{K}^{2}}{\sigma_m \, \omega_{K}}
\left( 1 - \cos \vartheta_{\bk, \bk'} \right).
\end{align}
Here $D_{O}$ is the optical phonon coupling constant, $\omega_{O}$ the optical phonon 
frequency, $D_{K}$ is the K-phonon coupling constant and $\omega_{K}$ the K-phonon frequency.

Regarding the scatterings of electrons with the substrate, the collision term $Q^{(el-sub)}$  is given by
\begin{equation*}
S^{(el-sub)}_{s', s}(\bk', \bk) = S^{(O-sub)}_{s', s}(\bk', \bk)+S^{(imp)}_{s', s}(\bk', \bk),
\end{equation*}
with
\begin{align*}
S^{(O-sub)}_{s', s}(\bk', \bk) = 
\left| G^{(O-sub)}(\bk', \bk) \right|^{2} & \left[ 
\left( n^{(O-sub)}_{\mathbf{q}} + 1 \right)
\delta \left( \eps_{s}(\bk) - \eps_{s'}(\bk') + \hbar \, 
\omega^{(O-sub)}_{\mathbf{q}}
\right) \right. \nonumber
\\
&
\left. + n^{(O-sub)}_{\mathbf{q}} \,
\delta \left( \eps_{s}(\bk) - \eps_{s'}(\bk') - \hbar \, 
\omega^{(O-sub)}_{\mathbf{q}}
\right) \right] ,
\end{align*}
where
\begin{align*}
&
\left| G^{(O-sub)}(\bk', \bk) \right|^{2} = \left| G^{(LO-sub)}(\bk', \bk) \right|^{2} + \left| G^{(TO-sub)}(\bk', \bk) \right|^{2} =
\dfrac{2}{(2 \, \pi)^{2}} \, \dfrac{\pi \, D_{O-sub}^{2}}{\sigma_m \, \omega_{O-sub}} .
\end{align*}
Concerning $S^{(imp)}$, we assume that the remote impurities interacting with the charge carriers are those located in a plane within the oxide at distance $d$ from the graphene sheet. The definition of the scattering rate for electron-impurity scattering is highly complex. Following \cite{Hwang2007}, we adopt the expression 
\begin{equation}\label{imp_scattering}
S^{(imp)}_{s', s}(\bk, \bk') = \dfrac{2 \pi}{\hbar}  \, \dfrac{n_{i}}{(2 \, \pi)^{2}} 
\left| \dfrac{V_{i}(|\bk - \bk'|, d)}{\epsilon(|\bk - \bk'|)} \right|^{2}
\dfrac{\left(1 + \cos \vartheta_{\bk, \bk'} \right)}{2}
\delta \left( \eps_s(\bk') - \eps_{s'}(\bk) \right) ,
\end{equation}
where, since the scattering is elastic, the only admissible cases are given by $s=s'$. In \eqref{imp_scattering} the parameter $n_{i}$ is the number of impurities per unit area while  $V_i$ is the screened Coulomb impurity potential given by
$$
V_{i}(|\bk - \bk'|, d) = 2 \, \pi e^{2} \, 
\dfrac{\exp(- \, d \, |\bk - \bk'|)}{\tilde{\kappa} \, |\bk - \bk'|},
$$
where $\dm \tilde{\kappa}$ is the effective dielectric constant, defined by $\dm 4 \pi  \epsilon_{0} \left( \kappa_{top} + \kappa_{bottom}  \right) / 2$, $\epsilon_{0}$ being the vacuum dielectric constant, $\kappa_{top}$ and $\kappa_{bottom}$ being the relative dielectric constants of the medium above and below the graphene layer respectively. The dielectric screening function $\epsilon$ in graphene  is approximated with the 2D finite temperature static random phase approximation (RPA)
\begin{equation*}
\epsilon\left(\left\vert \bk - \bk'\right\vert\right)=\left\lbrace
\begin{alignedat}{2}
&1 + \dfrac{q_{s}}{|\bk - \bk'|} - \dfrac{\pi \, q_{s}}{8 \, k_{F}}, && \qquad \mbox{if} \quad |\bk - \bk'| < 2 \, k_{F},\\
&1 + \dfrac{q_{s}}{|\bk - \bk'|} - \dfrac{q_{s} \sqrt{|\bk - \bk'|^{2} - 4 \, k_{F}^{2}}}{2 \, |\bk - \bk'|^{2}} - \dfrac{q_{s}}{4 \, k_{F}} \, \mbox{asin} \left( \dfrac{2 \, k_{F}}{|\bk - \bk'|} \right), && \qquad \mbox{otherwise},
\end{alignedat}
\right.
\end{equation*}
where $\dm q_{s} = \dfrac{4 \, e^{2} \, k_{F}}{\tilde{\kappa} \, \hbar \, v_{F}}$ is the effective Thomas-Fermi wave-vector for graphene with $\dm k_{F} = \dfrac{\varepsilon_F}{\hbar v_F}$  the Fermi wave-vector, $\varepsilon_F$ being the Fermi energy. Observe that if $\varepsilon_F=0$ the dielectric function is not defined. To overcome the problem, according to \cite{DasSarma} if $\vert\varepsilon_F\vert<0.04$ eV  we adopt the approximate expression $k_F=\sqrt{\frac{4\pi n}{g_s g_v}}$, being $n$ the 2D carrier density, $g_s=2$ and $g_v=2$ the spin and valley degeneracy. To simplify the computation, only for the evaluation of the scattering with impurities, the constant value $\varepsilon_F=$ 0.25 eV has been used (see Figure) which is the values of the work function. Indeed, in a situation which is spatially non homogeneous the Fermi level depends on the position but with good accuracy  its value ranges around the one given by the work function at the metallic contacts. See the Section \ref{SEC:Phys_sett} for further details.

We remark that there is a certain degree of uncertainty in the literature regarding the values of the parameters involved in the scattering processes. In our simulations, we use the physical parameters listed in Table~\ref{tabella} \cite{Hirai}. As demonstrated in \cite{CoMajRo}, the distance $d$ plays a critical role in accurately determining the electron velocity, and consequently, the electron mobility. In \cite{CoMajNaRo}, numerical simulations were carried out for the unipolar case, treating $d$ as a random variable. In this work we assume $d$ to be constant. This choice is supported by the findings in \cite{CoMajNaRo}, where the results obtained considering  $d$ a random variable were found to be practically equivalent to those using the average value of this parameter.

\begin{table}[ht]
\centering
\setlength{\tabcolsep}{8pt} 
\renewcommand{\arraystretch}{1.2} 
\begin{tabular}{lll}
\hline
Parameter & Description & Value \\
\hline
$v_{F}$ & Graphene Fermi velocity & $10^{8}$ cm/s \\
$\sigma_{m}$ & Graphene areal mass density & $7.6 \times 10^{-8}$ g/cm$^2$\\
$\hbar \, \omega_{O}$ & Optical phonon energy at the $\Gamma$ point & $164.6$ meV \\
$\hbar \, \omega_{K}$ & Optical phonon energy at the $K$ point & $124$ meV \\
$v_{p}$ & Acoustic phonon (sound) velocity & $2 \times 10^{6}$ cm/s \\
$D_{ac}$ & Acoustic deformation potential & $6.8$ eV \\
$D_{O}$ & Optical phonon deformation potential ($\Gamma$ point) & $10^{9}$ eV/cm \\
$D_{K}$ & Intervalley phonon deformation potential ($K$ point) & $3.5 \times 10^{8}$ eV/cm \\
$\hbar\omega_{O\text{-}sub}$ (SiO$_2$) & Surface optical phonon energy of the SiO$_2$ substrate & 55 meV \\
$D_{O\text{-}sub}$ (SiO$_2$) & Surface optical phonon coupling constant for SiO$_2$ & $5.14\times 10^7$ eV/cm \\
$n_i$ (SiO$_2$) & Charged impurity density in/on SiO$_2$ & $2.5 \times 10^{11}$ cm$^{-2}$ \\
$\kappa_{sub}$ (SiO$_2$) & Relative dielectric constant of SiO$_2$ & 3.9 \\
\hline
\end{tabular}

\caption{Physical parameters for the electron-phonon collision term.}
\label{tabella}
\end{table}

Once one gets the distribution functions $f_s$ solving the equations \eqref{EQ:Boltzmann},  the macroscopic quantities of interests are evaluated as averages with respect to the wave-vector. Of particular interest are 
the carrier density $n(t,\bx)$, the current density $\mathbf{j}^n(t,\bx)=\left(j^{n,x}(t,\bx),j^{n,y}(t,\bx)\right)$ and the energy density $\mathcal{E}^n (t,\bx)$. For the electron in the conduction band  they are defined as 
\begin{align*}
n(t,\bx) & = \frac{g_v g_s}{(2\pi)^2} \int_{\R^2} f_{+}(t,\bx,\bk) \, d\bk, \\
\mathbf{j}^n(t,\bx) & = -e\frac{g_v g_s}{(2\pi)^2}\int_{\R^2} \mathbf{v}_+(\bk) f_{+}(t,\bx,\bk) \, d\bk, \\
\mathcal{E}^n(t,\bx) & = \frac{g_v g_s}{(2\pi)^2}\int_{\R^2} \varepsilon_+(\bk) f_{+}(t,\bx,\bk) \, d\bk.
\end{align*}
All the quantities are evaluated at time $t$ and position $\bx$. Other macroscopic quantities directly related to the ones defined above are the electron mean velocity $\mathbf{V}^n(t,\bx)$ and the electron mean energy $E^n(t,\bx)$, which are defined as
\begin{align*}
\mathbf{V}^n(t,\bx) & = -\frac{1}{e}\frac{\mathbf{j}^n(t,\bx)}{n(t,\bx)},\\
E^n(t,\bx) & = \frac{\mathcal{E}^n(t,\bx)}{n(t,\bx)}.
\end{align*}
In the valence band, instead of electrons, it is convenient to consider holes \cite{BOOK:Jacoboni} to avoid integrability issues. It is possible to verify \cite{MajNaRo} that
\begin{equation}
f_h(t,\bx,\bk) = 1-f_-(t,\bx,\bk), \qquad \varepsilon_h(\bk)=\varepsilon_+(\bk), \qquad \mathbf{v}_h(\bk)=\mathbf{v}_+(\bk).
\end{equation}
where the subscript $h$ means quantity referred to holes. Similarly to electrons, the macroscopic quantities for holes are given by
\begin{align*}
p(t,\bx) & = \frac{g_v g_s}{(2\pi)^2} \int_{\R^2} f_{h}(t,\bx,\bk) \, d\bk, \\
\mathbf{j}^p(t,\bx) & = e\frac{g_v g_s}{(2\pi)^2}\int_{\R^2} \mathbf{v}_h(\bk) f_{h}(t,\bx,\bk) \, d\bk, \\
\mathcal{E}^p(t,\bx) & = \frac{g_v g_s,}{(2\pi)^2}\int_{\R^2} \varepsilon_h(\bk) f_{h}(t,\bx,\bk) \, d\bk, 
\end{align*}
where $p(t,\bx)$ is the hole density, $\mathbf{j}^p(t,\bx)=\left( j^{p,x}(t,\bx),j^{p,y}(t,\bx) \right)$ is the current density of holes and $\mathcal{E}^p$ the energy density of holes. The hole mean velocity $\mathbf{V}^p(t,\bx)$ and the hole mean energy $E^p(t,\bx)$ are defined as
\begin{align*}
\mathbf{V}^p(t,\bx) & = \frac{1}{e}\frac{\mathbf{j}^p(t,\bx)}{p(t,\bx)},\\
E^p(t,\bx) & = \frac{\mathcal{E}^p(t,\bx)}{p(t,\bx)}.
\end{align*}
The total current density is given by
\begin{equation}
\mathbf{j}^{tot}(t,\bx)=\mathbf{j}^n(t,\bx)-\mathbf{j}^p(t,\bx).
\end{equation}
Regarding the existence and uniqueness of solution  of the semiclassical Boltzmann equation \eqref{EQ:Boltzmann}, the results in \cite{Mus} can be adapted.

\section{Physical setting} \label{sec:setting}
\label{SEC:Phys_sett}
\begin{figure}[ht]
\centering
\includegraphics[width=0.75\textwidth]{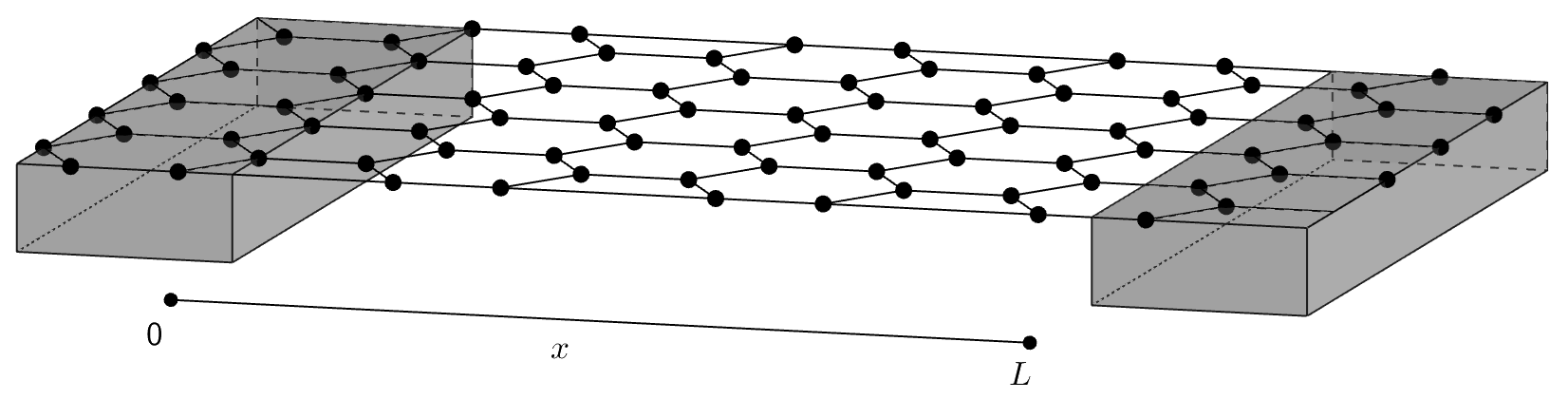}
\caption{Contacted monolayer graphene.}
\label{FIG:Cont_gr}
\end{figure}

As first test case, we consider a suspended monolayer graphene with two metallic contacts placed at the extremities, as in Fig. \ref{FIG:Cont_gr}. We take the electric field a constant external field. So the BTE is not coupled with the Poisson equation.

It is convenient to write  Eq.s \eqref{EQ:Boltzmann} in terms of energy $\varepsilon$ and angle $\theta$, measured from the transversal direction oriented as in Fig. \ref{FIG:Cont_gr}, instead of $\bk$. One has
\begin{equation}\label{k_rel}
(k_x,k_y)=\frac{\varepsilon}{\hbar v_F}(\cos\theta,\sin\theta),
\end{equation}
and the Jacobian of the transformation is $\eps/(\hbar v_F)^2$. In this way, the group velocity reads
\begin{equation}
\mathbf{v}_s=s v_F\frac{\bk}{\vert\bk\vert}=s v_F(\cos\theta,\sin\theta).
\end{equation}
Moreover,  we have
\begin{equation}
\nabla_{\bk} f_s = \hbar v_F \frac{\partial f_s}{\partial\varepsilon}\mathbf{e}_\varepsilon + \frac{\hbar v_F}{\varepsilon}\frac{\partial f_s}{\partial\theta}\mathbf{e}_\theta,
\end{equation}
where
\begin{equation*}
\begin{alignedat}{2}
\mathbf{e}_\varepsilon & = && \cos\theta \, \mathbf{i} + \sin\theta \, \mathbf{j},\\
\mathbf{e}_\theta & = -&&\sin\theta \, \mathbf{i} + \cos\theta \, \mathbf{j}.
\end{alignedat}
\end{equation*}


We consider the material along the directions of the contact as infinite. Therefore the symmetry of the structure allows us to treat the problem as one dimensional. 
If we denote with $x$ the abscissa in the direction orthogonal to the contacts the Boltzmann equations \eqref{EQ:Boltzmann} for the conduction band ($s =+$) and the valence band ($s =-$) write
\begin{equation}\label{EQ:Boltzmann_en}
\frac{\partial f_s}{\partial t} + s v_F\cos\theta\frac{\partial f_s}{\partial x} -e v_F E_x \cos\theta\frac{\partial f_s}{\partial\varepsilon} + \frac{e v_F E_x}{\varepsilon}\sin\theta\frac{\partial f_s}{\partial\theta} = Q(f_s,f_{-s}),
\end{equation}
where now $f_s=f_s(t,x,\varepsilon,\theta)$ with $t>0$, $x\in[0,L]$, $\varepsilon\in [0,+\infty[$ and $\theta\in[0,2\pi]$. The domain $[0,L]$, with $L>0$,  is the numerical domain representing the transversal section of the device. In order to keep the divergence form of the transport part of the equation, we multiply  Eq. \eqref{EQ:Boltzmann_en} by the Jacobian, obtaining
\begin{equation}
\frac{\eps}{(\hbar v_F)^2}\frac{\partial f_s}{\partial t} + s \frac{v_F}{(\hbar v_F)^2}\frac{\partial}{\partial x}\left(\eps\cos\theta f_s \right) - e E_x \frac{v_F}{(\hbar v_F)^2} \left[ \frac{\partial}{\partial\eps} \left( \eps\cos\theta f_s \right) - \frac{\partial}{\partial\theta} \left( \sin\theta f_s \right)\right] = \frac{\eps}{(\hbar v_F)^2} Q(f_s,f_{-s}).  \label{EQ:Boltzmann_div} 
\end{equation}

As initial conditions we assign the intrinsic equilibrium distribution
\begin{equation}\label{EQ:init_cond}
f_s(0,x,\varepsilon,\theta) = f^s_{FD}(\varepsilon,0), \qquad \forall x\in\left] 0, L \right[,
\end{equation}
where $f_{FD}$ is the Fermi-Dirac distribution
\begin{equation}
f^s_{FD}(\varepsilon,\varepsilon_F) = \frac{1}{1+\exp\left(\frac{s\varepsilon-\varepsilon_F}{k_B T}\right)},
\end{equation}
with $\varepsilon_F$ Fermi level.
At $x=0$ and $x=L$ we set  inflow boundary conditions
\begin{eqnarray}
& &f_s(t,0,\varepsilon,\theta) =  f^s_{FD}(\varepsilon,\eps_F), \qquad \forall t>0, \quad \forall \varepsilon \geq 0, \quad\forall \theta\in[-\pi/2,\pi/2],\\
& &f_s(t,L,\varepsilon,\theta)= f^s_{FD}(\varepsilon,\eps_F), \qquad \forall t>0, \quad \forall \varepsilon \geq 0, \quad\forall \theta\in[\pi/2,3/2\pi],
\end{eqnarray}
where $\eps_F$ takes an assigned value. 

\begin{figure}[ht]
\centering
\includegraphics[width=0.6\textwidth]{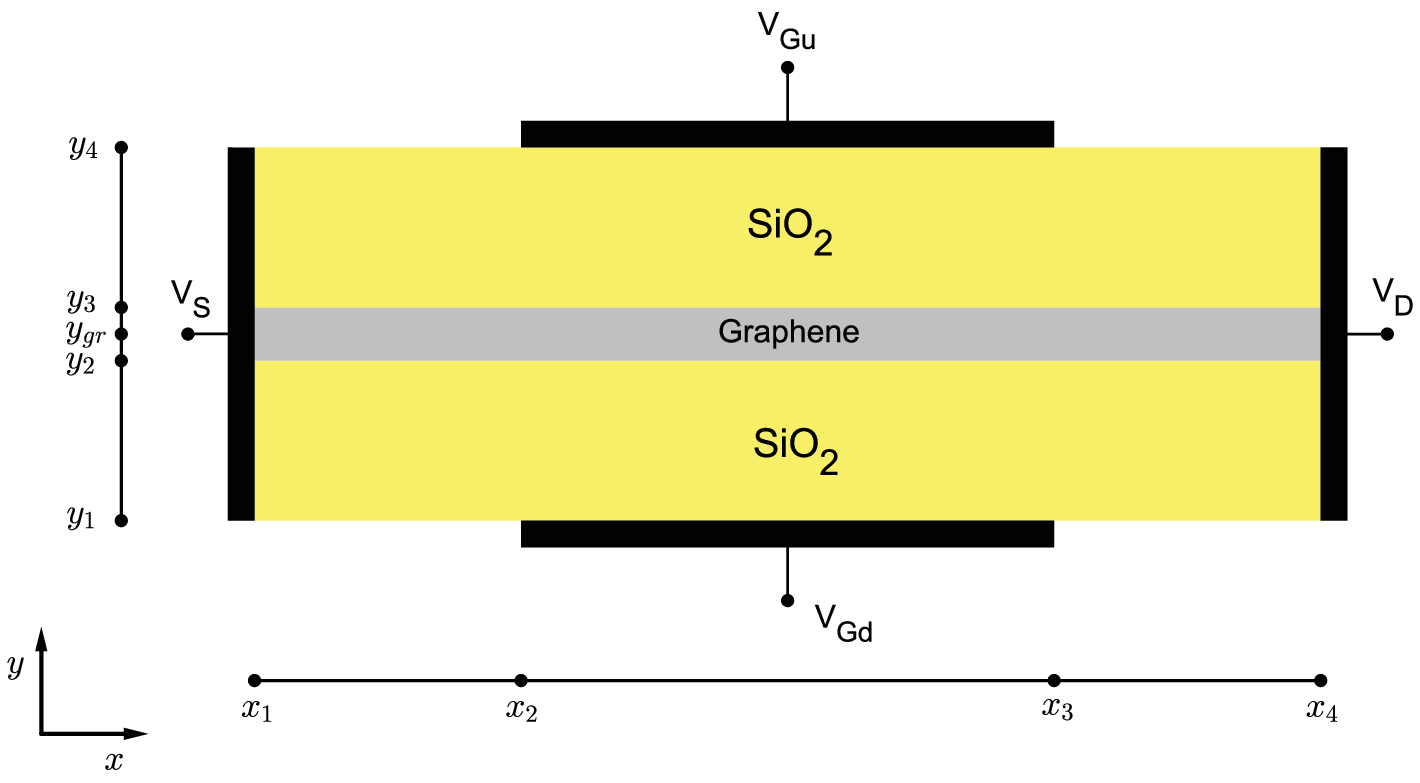}
\caption{GFET.}
\label{FIG:GFET}
\end{figure}

As a second test case, we simulate charge transport in a field-effect transistor (FET) having the active area made of monolayer large-area graphene (GFET). We choose the geometry introduced in \cite{NaRo_IEEE}, where  the same authors have performed the simulation by solving the drift-diffusion equations. The active zone, made of graphene, is placed between two strips of insulator, both of them being SiO$_2$. A schematic (cross-sectional view) of the configuration is shown in Fig. \ref{FIG:GFET}. The source and drain metallic contacts are directly attached to the graphene. The two gate contacts (up and down) are upon the oxide.

The domain consists of a rectangle of length $L$ and height $H$, which represents a section of the GFET, with $L=x_4-x_1$ and $H=y_4-y_1$. The Boltzmann equation \eqref{EQ:Boltzmann_en} is set for $x\in[x_1,x_4]$ along $y=y_{gr}$. We fix $x_1=y_1=0$. The initial condition is the same as that used in the previous case and is given by Eq. \eqref{EQ:init_cond}.

The boundary conditions need more explanation because modeling  the metal–graphene interface is a 
critical challenge. The contacts are treated as thermal reservoirs, wherein charge carriers are assumed to follow a Fermi–Dirac distribution. Charge injection across the interface is governed by the difference in work functions between the contact metal and graphene, which is inherently dependent on the specific material employed for the electrodes.
Copper is regarded as one of the most suitable substrates for graphene deposition via chemical vapor deposition (CVD), owing to its favorable surface properties. Notably, the interaction between copper and graphene is among the weakest observed for metallic contacts. First-principles calculations based on density functional theory (DFT) \cite{Chang} predict a Fermi level shift of    $\Delta \varepsilon_F = - $0.275 eV
 relative to the Dirac point at the Cu(111)/graphene junction, resulting in an induced n-type doping.
This theoretical prediction aligns with experimental observations, which report Fermi level shifts ranging from 0.20 eV \cite{FraVe} to 0.30 eV \cite{WaNi} in absolute magnitude. For subsequent analysis, we adopt a representative value of  $\Delta \varepsilon_F = -$ 0.25 eV
 corresponding to a built-in potential of $V_{built} = -  \Delta \varepsilon_F/ e$ = 0.25 V.
 Therefore  at $x=x_1$ and $x=x_4$  inflow boundary conditions
\begin{eqnarray}
& &f_s(t,0,\varepsilon,\theta) =  f^s_{FD}(\varepsilon,e\phi_W), \qquad \forall t>0, \quad \forall \varepsilon \geq 0, \quad\forall \theta\in[-\pi/2,\pi/2],\\
& &f_s(t,L,\varepsilon,\theta)= f^s_{FD}(\varepsilon,e\phi_W), \qquad \forall t>0, \quad \forall \varepsilon \geq 0, \quad\forall \theta\in[\pi/2,3/2\pi],
\end{eqnarray}
are assumed
where $\phi_W$ is the work function of the metallic contacts. We assume they are made of copper, so $e\phi_W=0.25$ eV.

For the complete simulation, a self consistent electric field is required. So we must couple the transport equation to Poisson equation \eqref{EQ:Poisson} for the electrostatic potential. In the specific case of the device of Fig. \ref{FIG:GFET}, we set
\begin{equation}
\epsilon(x,y)=\left\lbrace
\begin{alignedat}{2}
& \epsilon_{gr}, && \qquad \forall (x,y)\in[x_1,x_4]\times[y_2,y_3], \\
& \epsilon_{ox}, && \qquad \forall (x,y)\in[x_1,x_4]\times\left([y_1,y_2]\cup[y_3,y_4]\right),
\end{alignedat}
\right.
\end{equation}
where $\epsilon_{gr}=3.3\epsilon_0$ and $\epsilon_{ox}=3.9\epsilon_0$ are the dielectric constant of graphene and oxide (SiO$_2$) respectively, $\epsilon_0$ being the dielectric constant in the vacuum. The right hand side of eq. \eqref{EQ:Poisson} writes
\begin{equation}
h(x,y)=\left\lbrace
\begin{alignedat}{2}
& \frac{e(n(x)-p(x)-n_i)}{t_{gr}}, && \qquad \forall (x,y)\in[x_1,x_4]\times[y_2,y_3], \\
& 0, && \qquad \mbox{otherwise},
\end{alignedat}
\right.
\end{equation}
where $n(x)$ and $p(x)$ represent the electron and hole density respectively, $n_i=2.5\cdot 10^3$ $\mu$m$^{-2}$ is the areal density of the impurity charges at the graphene/oxide interface, and $t_{gr}=1$ nm is the thickness between the two strips of oxide, delimited by $y_2$ and $y_3$. We assume the charges in graphene are distributed in such a volume. For eq. \eqref{EQ:Poisson} we impose the following boundary conditions
\begin{equation}
\begin{alignedat}{2}
& \phi = V_D, && \qquad \forall (x,y)\in\left\lbrace x_1 \right\rbrace\times[y_1,y_4],\\
& \phi = V_S, && \qquad \forall (x,y)\in\left\lbrace x_4 \right\rbrace\times[y_1,y_4],\\
& \phi = V_{G_d}, && \qquad \forall (x,y)\in[x_2,x_3]\times\left\lbrace y_1 \right\rbrace,\\
& \phi = V_{G_u}, && \qquad \forall (x,y)\in[x_2,x_3]\times\left\lbrace y_4 \right\rbrace,\\
& \nabla_{\nu}\phi = 0, && \qquad \mbox{at the remaining part of the boundary},
\end{alignedat}
\end{equation}
where $V_S$ and $V_D$ are the source and drain voltages while $V_{G_d}$ and $V_{G_u}$ are the upper and down gate voltages. By using the gauge invariance of the electrostatic potential, we set $V_S = 0$ and $V_D = V_b$, $V_b$ being the bias voltage.

\section{Numerical method}\label{sec:numerics}
In this section the numerical scheme is presented. First we consider the Poisson equation and then the transport equation. 

\subsection{Discretization of the Poisson equation}
The Poisson equation is posed in the 2D domain $[0,L]\times[0,H]$. A mesh of $(N_x+1)\cdot (N_y+1)$ grid points is introduced and the domain $[0,L]\times[0,H]$ is partitioned in rectangular cells. The vertices of each cell are denoted by
\begin{equation*}
\begin{aligned}
x_{i + 1/2} & = i\Delta x, \qquad \mbox{for} \quad i = 0,1,\ldots, N_x,\\
y_{j + 1/2} & = j\Delta y, \qquad \mbox{for} \quad j = 0,1,\ldots, N_y,
\end{aligned}
\end{equation*}
where $\Delta x = L/N_x$ and $\Delta y = H/N_y$. The midpoints are given by
\begin{equation*}
\begin{aligned}
x_i & = \left(i-\frac{1}{2}\right)\Delta x, \qquad \mbox{for} \quad i = 1,2,\ldots, N_x,\\
y_j & = \left(j-\frac{1}{2}\right)\Delta y, \qquad \mbox{for} \quad j = 1,2,\ldots, N_y.
\end{aligned}
\end{equation*}
The reason for such a choice of the indices is related to the discretization of the Boltzmann equation.

We discretize the Poisson equation by standard finite differences. For the sake of simplifying the notation, the explicit dependence on time is omitted. Letting $\phi_{i+1/2,j+1/2}\approx \phi\left( x_{i+1/2},y_{j+1/2} \right)$, for $i = 0,1,\ldots, N_x$ and $j = 0,1,\ldots, N_y$, we get
\begin{equation}
\begin{aligned}
&\frac{\epsilon_{i+1,j+1/2}}{\Delta x^2}\phi_{i+3/2,j+1/2} - \frac{\epsilon_{i+1,j+1/2}+\epsilon_{i,j+1/2}}{\Delta x^2}\phi_{i+1/2,j+1/2} + \frac{\epsilon_{i,j+1/2}}{\Delta x^2}\phi_{i-1/2,j+1/2}\\
&+\frac{\epsilon_{i+1/2,j+1}}{\Delta y^2}\phi_{i+1/2,j+3/2} - \frac{\epsilon_{i+1/2,j+1}+\epsilon_{i+1/2,j}}{\Delta y^2}\phi_{i+1/2,j+1/2} + \frac{\epsilon_{i+1/2,j}}{\Delta y^2}\phi_{i+1/2,j-1/2} = h_{i+\frac{1}{2},j+\frac{1}{2}},
\end{aligned}
\end{equation}
where $\epsilon_{i,j+1/2}=\epsilon(x_i,y_{j+1/2})$ for all $i=1,2,\ldots,N_x$, $j=1,2,\ldots,N_y-1$, $\epsilon_{i+1/2,j}=\epsilon(x_{i+1/2},y_j)$ for all $i=1,2,\ldots,N_x-1$, $j=1,2,\ldots,N_y$, and $h_{i+\frac{1}{2},j+\frac{1}{2}}=h\left( x_{i+1/2},y_{j+1/2} \right)$ for all $i=1,2,\ldots,N_x-1$, $j=1,2,\ldots,N_y-1$. One gets $(N_x-1)(N_y-1)$ equations. Further $2(N_x+N_y)$ equations are determined by imposing the boundary conditions. We pay attention to choose $N_y$ such that $y_{gr}$ is a grid point.

\subsection{Discretization of the Boltzmann equation}
The Boltzmann equation for the distribution function $f(x,\varepsilon,\theta)$ is defined in the 1D set $[0,L]\times \left\{y_{gr} \right\}$ with respect to the spatial variable $x$ while the energy $\varepsilon$ and the angle $\theta$ vary in 
$[0,+\infty[\times[0,2\pi]$.

To discretize the semiclassical Boltzmann equations  a DG method is adopted. Since we expect an exponential decay of $f_+$ and $1- f_-$, as $\eps\to+\infty$ (see Theorem 1 in \cite{NaBoRo}), it is reasonable to choose a maximum value of energy $\eps_{max}$ and reduce the numerical domain in energy to a compact set. 

Practically the cutoff is determined through a worst-case numerical test: we considered a configuration with a constant high electric field, which continuously accelerates carriers toward higher energies, and zero-flux boundary conditions in momentum space. In this setting, the total carrier density should be conserved over time, since no particles are allowed to leave the computational domain. Therefore, any artificial loss of density can be directly attributed to an insufficient energy cutoff. The cutoff has been chosen sufficiently large so that density conservation is achieved numerically over the duration of the simulation.

\begin{figure}[ht]
\centering
\includegraphics[width=0.6\textwidth]{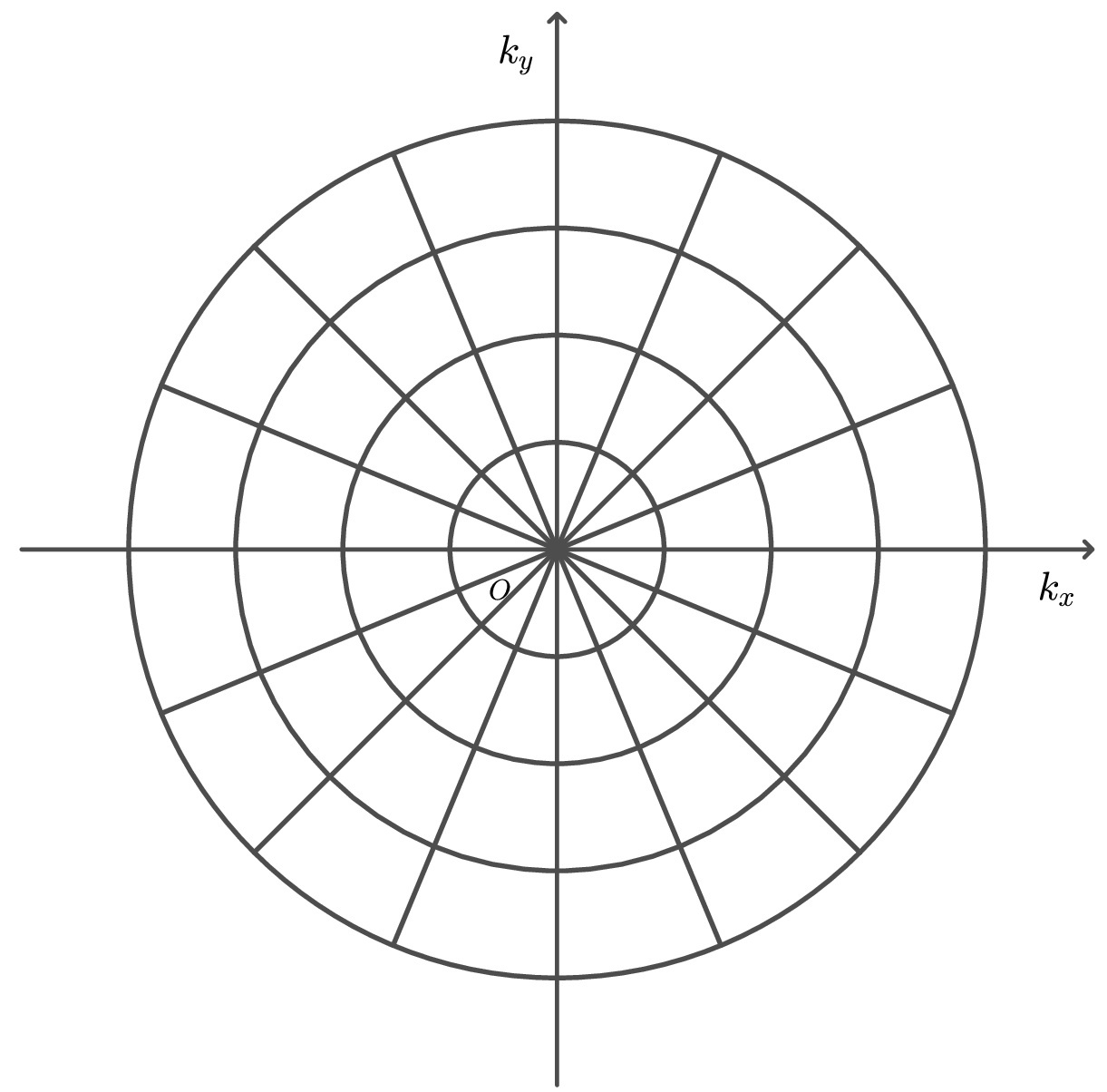}
\caption{Polar grid in the space of wave-vectors.}
\label{FIG:Pol_grid}
\end{figure}

Overall we have partitioned  the complete numerical domain  in $N_x\times N_\eps \times N_\theta$ open cells $C_{i,k,n}$, with $N_\eps$ and  $N_\theta$ positive integer, such that
$$
[0,L]\times[0,\eps_{max}]\times[0,2\pi] = \bigcup_{i=1}^{N_x} \bigcup_{k=1}^{N_\eps} \bigcup_{n=1}^{N_\theta} \overline{C}_{i,k,n},
$$
 $\overline{C}_{i,k,n}$ being the closure of $C_{i,k,n}$. We define three sets of indices
$$
I_x = \left\lbrace 1,2,\ldots,N_x \right\rbrace, \qquad I_\eps = \left\lbrace 1,2,\ldots,N_\eps \right\rbrace, \qquad I_\theta = \left\lbrace 1,2,\ldots,N_\theta \right\rbrace.
$$
For all $(i,k,n)\in I_x\times I_\eps\times I_\theta$ the cells are given by
$$
C_{i,k,n} = \left] x_{i-\frac{1}{2}},x_{i+\frac{1}{2}} \right[ \times \left\lbrace (\varepsilon,\theta)\in [0,\varepsilon_{max}]\times[0,2\pi] \, : \, \varepsilon_{k-\frac{1}{2}}<\varepsilon<\varepsilon_{k+\frac{1}{2}}, \, \theta_{n-\frac{1}{2}}<\theta<\theta_{n+\frac{1}{2}} \right\rbrace,
$$
where $x_{i\pm 1/2}$ are the same grid points of the discretization of Poisson's equation and
$$
\begin{aligned}
\eps_{k+1/2} & = k\Delta\eps, \qquad \mbox{for} \quad k = 0,1,\ldots,N_\eps,\\
\theta_{n+1/2} & = n\Delta\theta, \qquad \mbox{for} \quad n = 0,1,\ldots,N_\theta, 
\end{aligned}
$$
with $\Delta\eps=\eps_{max}/N_\eps$ and $\Delta\theta=2\pi/N_\theta$. The midpoints are given by
$$
\begin{aligned}
\eps_k & = \left( k-\frac{1}{2}\right)\Delta\eps, \qquad \mbox{for} \quad k = 1,2,\ldots,N_\eps,\\
\theta_n & = \left(n-\frac{1}{2}\right)\Delta\theta, \qquad \mbox{for} \quad n = 1,2,\ldots,N_\theta.
\end{aligned}
$$
Since the energy is proportional to the modulus of the wave-vector, the adopted grid is a polar one in the $\mathbf{k}$-space, as  shown in Fig. \ref{FIG:Pol_grid}.

We look for an approximation $\tilde{f}_s$ of $f_s$ such that, for each time $t\in[0,T]$, $\tilde{f}_s$ belongs to the finite dimensional space
$$
\tilde{V} = \left\lbrace v \in L^1\left( [0,L]\times[0,\eps_{max}]\times[0,2\pi] \right) \, : \, v\vert_{C_{i,k,n}}\in P^{m,r,l}(C_{i,k,n}) \quad \forall (i,k,n)\in I_x\times I_\eps\times I_\theta \right\rbrace,
$$
where $P^{m,r,l}(C_{i,k,n})$ denotes the space of the polynomials defined in $C_{i,k,n}$ of degree at most $m$ in $x$, at most $r$ in $\eps$ and at most $l$ in $\theta$. 

%

In order to determine the approximate solution $\tilde{f}_s$ we use a discretized weak formulation of our problem. 
In the spirit of the DG method, we require that in each cell for any test function $\tilde{v}\in \tilde{V}$ the approximation $\tilde{f}_s$ satisfies
\begin{equation}
\begin{aligned}
&\int_{C_{i,k,n}} \tilde{v}(x,\eps,\theta) \left\lbrace \frac{\eps}{(\hbar v_F)^2}\frac{\partial \tilde{f}_s}{\partial t} + s \frac{v_F}{(\hbar v_F)^2}\frac{\partial}{\partial x}\left(\eps\cos\theta \tilde{f}_s \right) - e E_x \frac{v_F}{(\hbar v_F)^2} \left[ \frac{\partial}{\partial\eps} \left( \eps\cos\theta \tilde{f}_s \right) - \frac{\partial}{\partial\theta} \left( \sin\theta \tilde{f}_s \right)\right] \right\rbrace \, dx \, d\eps \, d\theta \\
&= \int_{C_{i,k,n}} \tilde{v}(x,\eps,\theta) \frac{\eps}{(\hbar v_F)^2} Q(\tilde{f}_s,\tilde{f}_{-s}) \, dx \, d\eps \, d\theta.
\end{aligned} \label{EQ:weak_approx}
\end{equation}
Observe that, at variance with finite element approach,  in general $\tilde{f}_s$ is allowed to be  discontinuous along the boundary of adjacent cells.

We choose a piece-wise linear approximation in $x$ and a piece-wise constant approximation in $\eps$ and $\theta$
\begin{equation}\label{EQ:FF_approx}
\tilde{f}_s(t,x,\eps,\theta) = a_{i,k,n}^s(t) + \frac{2(x-x_i)}{\Delta x} b_{i,k,n}^s(t), \qquad \forall(x,\eps,\theta)\in C_{i,k,n}.
\end{equation}
An orthogonal basis for the polynomial space $P^{1,0,0}(C_{i,k,n})$ is
\begin{equation}
\tilde{v}^0(x,\eps,\theta)=1, \qquad \tilde{v}^1(x,\eps,\theta)=\frac{2(x-x_i)}{\Delta x}.
\end{equation}
The reconstruction of the approximation of $f_s$ in all the computational domain is
\begin{equation}
\tilde{f}_s = \sum_{i=1}^{N_x}\sum_{k=1}^{N_\eps}\sum_{n=1}^{N_\theta} \left[ a_{i,k,n}^s(t) + \frac{2(x-x_i)}{\Delta x} b_{i,k,n}^s(t) \right] \chi_{C_{i,k,n}}(x,\eps,\theta),
\end{equation}
where $\chi_{C_{i,k,n}}(\cdot,\cdot,\cdot)$ represents the characteristic function of $C_{i,k,n}$.

Now we project the equations in the finite dimensional space $\tilde{V}$ by substituting the basis elements $\tilde{v}^0$ and $\tilde{v}^1$
to $\tilde{v}$ in the approximate weak formulation \eqref{EQ:weak_approx}. For all $i = 1, 2, \ldots ,N_x$, for all $k=1,2,\ldots,N_\eps$ and for all $n=1,2,\ldots,N_\theta$ we discretize each term of equation separately.
 
\subsubsection{Discretization of the free-streaming case}
The terms involving the time derivatives become
\begin{equation}
\int_{C_{i,k,n}} \frac{\eps}{(\hbar v_F)^2}\frac{\partial \tilde{f}_s}{\partial t} \, dx \, d\eps \, d\theta = \Delta x \Delta \theta \frac{\eps_{k+1/2}^2-\eps_{k-1/2}^2}{2(\hbar v_F)^2} \frac{d}{dt} a_{i,k,n}^s(t),
\end{equation}
if we choose $\tilde{v}^0$ as test function, while they become
\begin{equation}
\begin{aligned}
&\int_{C_{i,k,n}} \frac{2(x-x_i)}{\Delta x} \frac{\eps}{(\hbar v_F)^2}\frac{\partial \tilde{f}_s}{\partial t} \, dx \, d\eps \, d\theta \\
& = \Delta\theta\frac{\eps_{k+1/2}^2-\eps_{k-1/2}^2}{2(\hbar v_F)^2} \left[ \frac{d}{dt} a_{i,k,n}^s(t)\int_{x_{i-1/2}}^{x_{i+1/2}} \frac{2(x-x_i)}{\Delta x} \, dx + \frac{d}{dt} b_{i,k,n}^s(t)\int_{x_{i-1/2}}^{x_{i+1/2}} \frac{4(x-x_i)^2}{(\Delta x)^2} \, dx  \right]\\
& = \frac{\Delta x}{3}\Delta\theta\frac{\eps_{k+1/2}^2-\eps_{k-1/2}^2}{2(\hbar v_F)^2} \frac{d}{dt} b_{i,k,n}^s(t),
\end{aligned}
\end{equation}
if we choose $\tilde{v}^1$ as test function.

\begin{figure}[ht]
\centering
\includegraphics[width=0.9\textwidth]{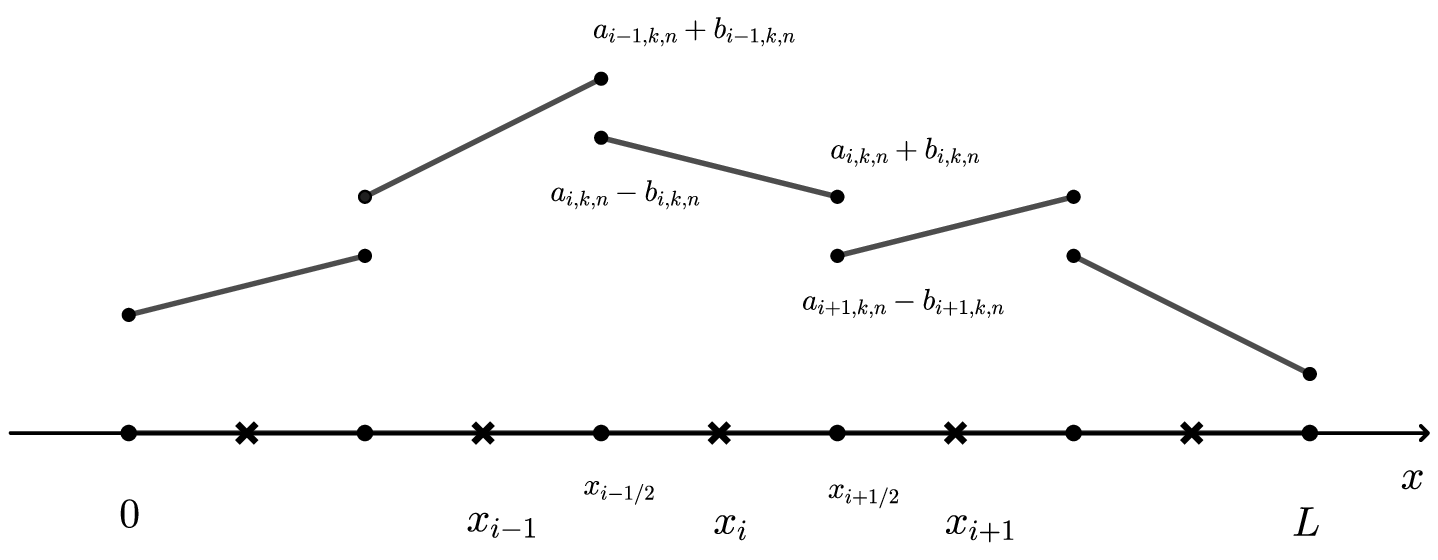}
\caption{Discretization with respect to $x$ where $\eps$ and $\theta$ are fixed.}
\label{FIG:Disc_flux_x}
\end{figure}

The terms involving the space derivatives, for $m=0,1$ can be written as
\begin{equation}
\begin{aligned}
&\int_{C_{i,k,n}} \tilde{v}^m \left\lbrace s\frac{v_F}{(\hbar v_F)^2}\eps\cos\theta \frac{\partial \tilde{f}_s}{\partial x} \right\rbrace \, dx \, d\eps \, d\theta  \\
& = s\frac{v_F}{(\hbar v_F)^2} \int_{\eps_{k-1/2}}^{\eps_{k+1/2}} \int_{\theta_{n-1/2}}^{\theta_{n+1/2}} \int_{x_{i-1/2}}^{x_{i+1/2}} \left[ \frac{\partial}{\partial x} \left( \tilde{v}^m \eps\cos\theta \tilde{f}_s \right) - \frac{\partial \tilde{v}^m}{\partial x} \eps\cos\theta \tilde{f}_s \right] \, dx \, d\theta \, d\eps .
\end{aligned}
\end{equation}
In particular, for $\tilde{v}^0$, we have
\begin{equation}\label{EQ:der_x_v_0}
\begin{aligned}
&\int_{C_{i,k,n}} s\frac{v_F}{(\hbar v_F)^2}\eps\cos\theta \frac{\partial \tilde{f}_s}{\partial x} \, dx \, d\eps \, d\theta \\
& = s\frac{v_F}{(\hbar v_F)^2} \int_{\eps_{k-1/2}}^{\eps_{k+1/2}} \int_{\theta_{n-1/2}}^{\theta_{n+1/2}} \eps\cos\theta\left[  \tilde{f}_s(t,x_{i+1/2},\eps,\theta) - \tilde{f}_s(t,x_{i-1/2},\eps,\theta) \right] \, d\theta \, d\eps,
\end{aligned}
\end{equation}
and, for $\tilde{v}^1$, we have
\begin{equation}\label{EQ:der_x_v_1}
\begin{aligned}
&\int_{C_{i,k,n}} s\frac{v_F}{(\hbar v_F)^2}\frac{2(x-x_i)}{\Delta x}\eps\cos\theta \frac{\partial f_s}{\partial x} \, dx \, d\eps \, d\theta \\
& = s\frac{v_F}{(\hbar v_F)^2} \int_{\eps_{k-1/2}}^{\eps_{k+1/2}} \int_{\theta_{n-1/2}}^{\theta_{n+1/2}} \eps\cos\theta\left[  \tilde{f}_s(t,x_{i+1/2},\eps,\theta) + \tilde{f}_s(t,x_{i-1/2},\eps,\theta) - 2 a_{i,k,n}^s(t) \right] \, d\theta \, d\eps.
\end{aligned}
\end{equation}
Since $\tilde{f}_s$ is discontinuous on the boundaries of the cells we must replace the flux $\tilde{f}_s(t,x_{i+1/2}, \eps, \theta)$ by a numerical flux depending on the two values of $\tilde{f}_s$ along $\left\lbrace x_{i+1/2} \right\rbrace \times \left] \eps_{k-1/2},\eps_{k+1/2} \right[\times\left] \theta_{n-1/2},\theta_{n+1/2} \right[$, that is, for all $(\eps,\theta)\in \left] \eps_{k-1/2},\eps_{k+1/2} \right[\times\left] \theta_{n-1/2},\theta_{n+1/2} \right[$
$$
\tilde{f}_s(t,x_{i+1/2}, \eps, \theta) = F^s_{i+1/2,k,n}\left( \tilde{f}_s(t,x_{i+1/2}^-, \eps, \theta),\tilde{f}_s(t,x_{i+1/2}^+, \eps, \theta) \right),
$$
where (see Fig. \ref{FIG:Disc_flux_x})
\begin{equation}
\begin{aligned}
\tilde{f}_s(t,x_{i+1/2}^-, \eps, \theta) & \approx a^s_{i,k,n}(t) + b^s_{i,k,n}(t)=:\tilde{f}^{s,-}_{i+1/2,k,n}(t),\\
\tilde{f}_s(t,x_{i+1/2}^+, \eps, \theta) & \approx a^s_{i+1,k,n}(t) - b^s_{i+1,k,n}(t)=:\tilde{f}^{s,+}_{i+1/2,k,n}(t),
\end{aligned}
\end{equation}
and $F^s_{i+1/2,k,n}$ a suitable numerical recipe.

Therefore, eq. \eqref{EQ:der_x_v_0} becomes
\begin{equation}
\int_{C_{i,k,n}} s\frac{v_F}{(\hbar v_F)^2}\eps\cos\theta \frac{\partial \tilde{f}_s}{\partial x} \, dx \, d\eps \, d\theta = \left[  F^s_{i+1/2,k,n}(t) - F^s_{i-1/2,k,n}(t) \right] M^s_{k,n},
\end{equation}
while eq. \eqref{EQ:der_x_v_1} reads
\begin{equation}
\int_{C_{i,k,n}} s\frac{v_F}{(\hbar v_F)^2}\eps\cos\theta \frac{\partial \tilde{f}_s}{\partial x} \, dx \, d\eps \, d\theta = \left[  F^s_{i+1/2,k,n}(t) + F^s_{i-1/2,k,n}(t) -2 a^s_{i,k,n}(t) \right] M^s_{k,n},
\end{equation}
where we have set
\begin{equation}
M^s_{k,n} := s\frac{v_F}{(\hbar v_F)^2} \int_{\eps_{k-1/2}}^{\eps_{k+1/2}} \int_{\theta_{n-1/2}}^{\theta_{n+1/2}} \eps\cos\theta \, d\theta \, d\eps = s\frac{v_F}{(\hbar v_F)^2} \frac{\eps^2_{k+1/2} - \eps^2_{k-1/2}}{2}\left(\sin \theta_{n+1/2} - \sin \theta_{n-1/2}\right).
\end{equation}
We remark that the reconstructions at $x_{1/2}=0$ and $x_{N_x+1/2}=L$ are given by the boundary conditions. Since the equation is linear with respect to $x$, as numerical flux function $F^s_{i+1/2,k,n}$ we adopt
\begin{equation}\label{EQ:num_flux_x}
F^s_{i+1/2,k,n}\left(\tilde{f}^{s,-}_{i+1/2,k,n},\tilde{f}^{s,+}_{i+1/2,k,n}\right) = \frac{\tilde{f}^{s,-}_{i+1/2,k,n}+\tilde{f}^{s,+}_{i+1/2,k,n}}{2}-\frac{1-\eta}{2}\,\mathrm{sign}(M^s_{k,n})\left(\tilde{f}^{s,-}_{i+1/2,k,n}-\tilde{f}^{s,+}_{i+1/2,k,n}\right),
\end{equation}
where $0\leq\eta\leq 1$ for stability \cite{BOOK:Hesthaven}. Observe that $\eta=1$ returns the central flux and $\eta=0$ the upwind one. The most appropriate value depends on the specific equation and in many cases it must be determined numerically. In the present paper we adopt the upwind flux.
\subsubsection{Discretization of the drift term}
\begin{figure}[ht]
\centering
\includegraphics[width=0.5\textwidth]{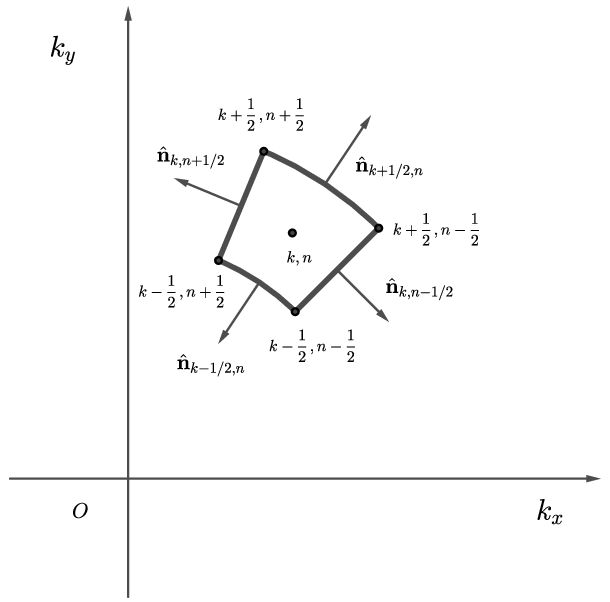}
\caption{Representation of a generic cell for the computation of the numerical fluxes.}
\label{FIG:Pol_cell}
\end{figure}

In order to approximate the drift term, we first consider a piece-wise linear interpolation of the electrostatic potential $\phi$, that is
\begin{equation}
\phi(x,y_{gr}) \approx \phi_{i-1/2,j_{gr}} + \frac{x-x_{i-1/2}}{\Delta x} \left( \phi_{i+1/2,j_{gr}} - \phi_{i-1/2,j_{gr}} \right) , \qquad \forall x \in \left[ x_{i-1/2}, x_{i+1/2} \right],
\end{equation}
where $j_{gr}$ is the index of the discretization corresponding to $y_{gr}$. Consequently, the approximation of the electrical field $E_x$ reads
\begin{equation}
E_x(x,y_{gr}) = - \frac{\partial \phi(x,y_{gr})}{\partial x} \approx  - \frac{\phi_{i+1/2,j_{gr}} - \phi_{i-1/2,j_{gr}}}{\Delta x}, \qquad \forall x \in \left] x_{i-1/2}, x_{i+1/2} \right[.
\end{equation}
Therefore, we set
\begin{equation}
E_x^i = - \frac{\phi_{i+1/2,j_{gr}} - \phi_{i-1/2,j_{gr}}}{\Delta x}, \qquad \forall x \in \left] x_{i-1/2}, x_{i+1/2} \right[.
\end{equation}
Overall, the approximation of the drift term reads
\begin{equation}\label{EQ:transport_term}
\begin{aligned}
& \int_{C_{i,k,n}} \tilde{v}^m \left\lbrace -e E_x \frac{v_F}{(\hbar v_F)^2}\left[ \frac{\partial}{\partial\eps}(\eps\cos\theta \tilde{f}_s) - \frac{\partial}{\partial\theta}(\sin\theta \tilde{f}_s) \right] \right\rbrace \, dx \, d\eps \, d\theta\\
& \approx - \frac{eE_x^i v_F}{(\hbar v_F)^2} \int_{x_i-1/2}^{x_i+1/2} \tilde{v}^m \left\lbrace \int_{\theta_{n-1/2}}^{\theta_{n+1/2}} \left[ \eps\cos\theta \tilde{f}_s \right]_{\eps_{k-1/2}}^{\eps_{k+1/2}} \, d\theta - \int_{\eps_{k-1/2}}^{\eps_{k+1/2}} \left[ \sin\theta \tilde{f}_s \right]_{\theta_{n-1/2}}^{\theta_{n+1/2}} d\eps \right\rbrace \, dx.
\end{aligned}
\end{equation}
On account of the fact that $ \tilde{f}_s$ is discontinuous on the boundary of the cell,  in $\eps=\eps_{k\pm 1/2}$ and $\theta=\theta_{n\pm 1/2}$ we have not a single value, so we have to introduce a numerical flux function on the boundary of the cell. Since we adopt a piece-wise constant approximation with respect to $\eps$ and $\theta$, the adoption of a numerical flux function reflects on the degrees of freedom of the DG approximation directly. In particular, we take
\begin{equation}
\begin{aligned}
 \tilde{f}_s(t,x,\varepsilon_{k\pm 1/2},\theta) & = a_{i,k\pm 1/2,n}^s + \frac{2(x-x_i)}{\Delta x}b_{i,k\pm 1/2,n}^s \qquad \forall x \in \left] x_{i-1/2}, x_{i+1/2} \right[, \quad \forall \theta\in \left] \theta_{n-1/2},\theta_{n+1/2} \right[,\\
 \tilde{f}_s(t,x,\varepsilon,\theta_{n\pm 1/2}) & = a_{i,k,n\pm 1/2}^s + \frac{2(x-x_i)}{\Delta x}b_{i,k,n\pm 1/2}^s \qquad \forall x \in \left] x_{i-1/2}, x_{i+1/2} \right[, \quad \forall \eps\in \left] \eps_{k-1/2},\eps_{k+1/2} \right[.\\
\end{aligned}
\end{equation}
Therefore, for $m=0$, the right hand side of  equation \eqref{EQ:transport_term} becomes
\begin{equation}\label{EQ:transport_term_0}
\begin{aligned}
& - \frac{eE_x^i v_F}{(\hbar v_F)^2} \Delta x\left\lbrace \left[ \eps_{k+1/2} \, a^s_{i,k+1/2,n}-\eps_{k-1/2} \, a^s_{i,k-1/2,n} \right] 
\left( \sin\theta_{n+1/2} -  \sin\theta_{n-1/2} \right) \right.\\
&\left. - \left[ \sin\theta_{n+1/2} \, a^s_{i,k,n+\frac{1}{2}}-\sin\theta_{n-1/2} \, a^s_{i,k,n-1/2} \right] \Delta\eps \right\rbrace,
\end{aligned}
\end{equation}
and, for $m=1$, it reads
\begin{equation}\label{EQ:transport_term_1}
\begin{aligned}
& - \frac{eE_x^i v_F}{(\hbar v_F)^2} \frac{\Delta x}{3}\left\lbrace \left[ \eps_{k+1/2} \, b^s_{i,k+1/2,n}-\eps_{k-1/2} \, b^s_{i,k-1/2,n} \right] 
\left( \sin\theta_{n+1/2} -  \sin\theta_{n-1/2} \right) \right.\\
&\left. - \left[ \sin\theta_{n+1/2} \, b^s_{i,k,n+\frac{1}{2}}-\sin\theta_{n-1/2} \, b^s_{i,k,n-1/2} \right] \Delta\eps \right\rbrace.
\end{aligned}
\end{equation}
The quantities $a^s_{i,k\pm 1/2,n}$, $a^s_{i,k,n\pm 1/2}$, $b^s_{i,k\pm 1/2,n}$, and $b^s_{i,k,n\pm 1/2}$ represent the numerical flux functions and depend on  the nearest neighbor degrees of freedom. For a linear equation, the simplest choice is to adopt the numerical flux \eqref{EQ:num_flux_x}. Following \cite{CoMajRo}, we assume, instead, a uniformly non-oscillatory (UNO) piece-wise linear reconstruction. For a generic 1D situation, such an approximation of a function $g(z)$ in $z_{j+1/2}$ is obtained by a numerical flux function $g_{j+1/2}=g_{j+1/2}\left(g_{j-1},g_{j},g_{j+1},g_{j+2}\right)$, where $g_j$ indicates the cell average of $g$ in $\left] z_j-\frac{\Delta z}{2},z_j+\frac{\Delta z}{2} \right[$. Once identified the wind velocity $w$, by approximating with the first term of the Taylor expansion, we have
\begin{equation}
g_{j+1/2}\approx\left\lbrace
\begin{alignedat}{2}
& g_j + \frac{\Delta z}{2}g_j', && \qquad \mbox{if} \quad w > 0,\\
& g_{j+1} - \frac{\Delta z}{2}g_{j+1}', && \qquad \mbox{if} \quad w<0,
\end{alignedat}
\right.
\end{equation}
where 
\begin{equation}
g'_j = \mathrm{MinMod}\left( \frac{g_j-g_{j-1}}{\Delta z},\frac{g_{j+1}-g_{j}}{\Delta z} \right),
\end{equation}
with
\begin{equation}
\mathrm{MinMod}(\alpha,\beta) = \left\lbrace
\begin{alignedat}{2}
& \min\left( \vert \alpha \vert, \vert \beta \vert \right)\mathrm{sgn}(\alpha), && \qquad \mbox{if} \quad \alpha\beta >0,\\
& 0, && \qquad \mbox{otherwise}.
\end{alignedat}
\right.
\end{equation}
In the case of equations \eqref{EQ:transport_term_0} and \eqref{EQ:transport_term_1}, we assume as wind velocity related to $a_{i,k+1/2,n}^s$ and $b_{i,k+1/2,n}^s$ the quantity
\begin{equation}
w_{i,k+1/2,n} = \mathrm{sgn}\left(-E_x^i\left(\mathbf{i}\cdot\hat{\mathbf{n}}_{k+1/2,n} \right) \left( \sin\theta_{n+1/2} -  \sin\theta_{n-1/2} \right)\right),
\end{equation}
and, as wind velocity related to $a_{i,k,n+1/2}^s$ and $b_{i,k,n+1/2}^s$ the quantity
\begin{equation}
w_{i,k,n+1/2} = \mathrm{sgn}\left( \Delta\eps \, E_x^i\left(\mathbf{i}\cdot\hat{\mathbf{n}}_{k,n+1/2} \right)\right),
\end{equation}
where $\hat{\mathbf{n}}_{k+1/2,n}$ is the outer normal direction of the edge of the cell $C_{i,k,n}$ identified by $\eps=\eps_{k+1/2}$, and $\hat{\mathbf{n}}_{k,n+1/2}$ is the outer normal direction of the edge of the cell $C_{i,k,n}$ identified by $\theta=\theta_{n+1/2}$, see Fig. \ref{FIG:Pol_cell}. Since the above discretization holds for each $k=0,1,\ldots,N_\eps$ and for each $n=0,1,\ldots,N_\theta$, we need to extend the grid with two ghosts cells to the ends of both the radial and the angular direction. Therefore, in addition, we assume zero radial inflow at $\eps=0$ and zero radial outflow at $\eps=\eps_{max}$, while we assume periodic boundary conditions with respect to the angle. The choice described above assures a higher accuracy with respect to the flux \eqref{EQ:num_flux_x} and the total variation diminishing (TVD) property \cite{HartenOsher}. Although it is desirable for accuracy to adopt a piece-wise DG discretization also with respect to the energy and the angle, our choice aims to maintain a moderate computational costs.
\subsubsection{Discretization of the collision term}
To approximate the collision term, and to avoid cumbersome notation, we first note that it can be expressed as
\begin{equation}
\begin{aligned}
Q(f_s,f_{-s})  = \sum_{s'} \sum_{\lambda} \int_0^{+\infty} d\eps' \int_0^{2\pi} d\theta' \frac{\eps'}{(\hbar v_F)^2} &\left[ S^{(\lambda)}_{s',s}(\eps',\theta',\eps,\theta)f_{s'}(t,x,\eps',\theta')(1-f_{s}(t,x,\eps,\theta))\right.\\
&-\left. S^{(\lambda)}_{s,s'}(\eps,\theta,\eps',\theta')f_{s}(t,x,\eps,\theta)(1-f_{s'}(t,x,\eps',\theta')) \right].
\end{aligned}
\end{equation}
where the index $\lambda$ runs over the possible types of scattering.

We set
\begin{align*}
&A_{s',s}^{\substack{k',n'\\k,n}} = \frac{1}{(\hbar v_F)^4} \int_{\varepsilon_{k-\frac{1}{2}}}^{\varepsilon_{k+\frac{1}{2}}} d\varepsilon \int_{\theta_{n-\frac{1}{2}}}^{\theta_{n+\frac{1}{2}}} d\theta \int_{\varepsilon_{k'-\frac{1}{2}}}^{\varepsilon_{k'+\frac{1}{2}}} d\varepsilon' \int_{\theta_{n'-\frac{1}{2}}}^{\theta_{n'+\frac{1}{2}}} d\theta' \, \varepsilon'\varepsilon\sum_{\lambda} \left[ D^{(\lambda)}+E^{(\lambda)}\cos(\theta-\theta') \right]\\
&\times \left[ (n_{\mathbf{q}}^{(\lambda)}+1)\delta(s'\varepsilon'-s\varepsilon+\hbar\omega^{(\lambda)})+n_{\mathbf{q}}^{(\lambda)}\delta(s'\varepsilon'-s\varepsilon-\hbar\omega^{(\lambda)}) \right]\\
&= \frac{1}{(\hbar v_F)^4} \sum_{\lambda} \int_{\theta_{n-\frac{1}{2}}}^{\theta_{n+\frac{1}{2}}} d\theta \int_{\theta_{n'-\frac{1}{2}}}^{\theta_{n'+\frac{1}{2}}} d\theta' \left[ D^{(\lambda)}+E^{(\lambda)}\cos(\theta-\theta') \right] \\
& \times \int_{\varepsilon_{k-\frac{1}{2}}}^{\varepsilon_{k+\frac{1}{2}}} d\varepsilon \int_{\varepsilon_{k'-\frac{1}{2}}}^{\varepsilon_{k'+\frac{1}{2}}} d\varepsilon' \,\varepsilon'\varepsilon \left[ (n_{\mathbf{q}}^{(\lambda)}+1)\delta\left(\varepsilon'-\frac{s}{s'}\varepsilon+\frac{\hbar\omega^{(\lambda)}}{s'}\right)+n_{\mathbf{q}}^{(\lambda)}\delta\left(\varepsilon'-\frac{s}{s'}\varepsilon-\frac{\hbar\omega^{(\lambda)}}{s'}\right) \right],
\end{align*}
where the coefficients $D^{(\lambda)}$ and $E^{(\lambda)}$ are related to the scattering rates.

We have
\begin{align*}
&\int_{\theta_{n-\frac{1}{2}}}^{\theta_{n+\frac{1}{2}}} d\theta \int_{\theta_{n'-\frac{1}{2}}}^{\theta_{n'+\frac{1}{2}}} d\theta' \left[ D^{(\lambda)}+E^{(\lambda)}\cos(\theta-\theta') \right] \\
& = D^{(\lambda)} \Delta\theta^2 + 4 E^{(\lambda)} \sin^2\left( \frac{\Delta\theta}{2} \right)\cos\left( \frac{\theta_{n-\frac{1}{2}}+\theta_{n+\frac{1}{2}}}{2}-\frac{\theta_{n'-\frac{1}{2}}+\theta_{n'+\frac{1}{2}}}{2} \right)
\end{align*}
and
\begin{align*}
&\int_{\varepsilon_{k-\frac{1}{2}}}^{\varepsilon_{k+\frac{1}{2}}} d\varepsilon \int_\mathbb{R} d\varepsilon' \, \chi_{\left[\varepsilon_{k'-\frac{1}{2}},\varepsilon_{k'+\frac{1}{2}}\right]}(\varepsilon') \varepsilon'\varepsilon \left[ (n_{\mathbf{q}}^{(\lambda)}+1)\delta\left(\varepsilon'-\frac{s}{s'}\varepsilon+\frac{\hbar\omega^{(\lambda)}}{s'}\right)+n_{\mathbf{q}}^{(\lambda)}\delta\left(\varepsilon'-\frac{s}{s'}\varepsilon-\frac{\hbar\omega^{(\lambda)}}{s'}\right) \right]\\
&=\int_{\varepsilon_{k-\frac{1}{2}}}^{\varepsilon_{k+\frac{1}{2}}} d\varepsilon \left\lbrace (n_{\mathbf{q}}^{(\lambda)}+1)\varepsilon\left( \frac{s}{s'}\varepsilon-\frac{\hbar\omega^{(\lambda)}}{s'} \right)\chi_{\left[\varepsilon_{k'-\frac{1}{2}},\varepsilon_{k'+\frac{1}{2}}\right]}\left( \frac{s}{s'}\varepsilon-\frac{\hbar\omega^{(\lambda)}}{s'} \right) \right.\\
&\left. n_{\mathbf{q}}^{(\lambda)}\varepsilon\left( \frac{s}{s'}\varepsilon+\frac{\hbar\omega^{(\lambda)}}{s'} \right)\chi_{\left[\varepsilon_{k'-\frac{1}{2}},\varepsilon_{k'+\frac{1}{2}}\right]}\left( \frac{s}{s'}\varepsilon+\frac{\hbar\omega^{(\lambda)}}{s'} \right) \right\rbrace\\
&=(n_{\mathbf{q}}^{(\lambda)}+1)\left[ \frac{s}{s'}\frac{\varepsilon^3}{3}-\frac{\hbar\omega^{(\lambda)}}{s'}\frac{\varepsilon^2}{2} \right]_{\min I^-}^{\max I^-}+n_{\mathbf{q}}^{(\lambda)}\left[ \frac{s}{s'}\frac{\varepsilon^3}{3}+\frac{\hbar\omega^{(\lambda)}}{s'}\frac{\varepsilon^2}{2} \right]_{\min I^+}^{\max I^+},
\end{align*}
where
\begin{equation*}
I^\pm = \left\lbrace \varepsilon\in\mathbb{R}_0^+ \, : \, \varepsilon_{k-\frac{1}{2}} \leq \varepsilon \leq \varepsilon_{k+\frac{1}{2}}, \,\, \varepsilon_{k'-\frac{1}{2}} \leq \frac{s}{s'}\varepsilon\pm\frac{\hbar\omega^{(\lambda)}}{s'} \leq \varepsilon_{k'+\frac{1}{2}} \right\rbrace.
\end{equation*}
Therefore, the approximation of the collision term reads
\begingroup
\allowdisplaybreaks
\begin{align*}
&\int_{C_{i,k,n}}  \tilde{v}^m\frac{\eps}{(\hbar v_F)^2} Q( \tilde{f}_s, \tilde{f}_{-s}) \, dx \, d\eps \, d\theta = \sum_{s'} \sum_{k'=1}^{N_\eps} \sum_{n'=1}^{N_\theta} \int_{x_{i-1/2}}^{x_{i+1/2}} dx \,  \tilde{v}^m \left[ A_{s',s}^{\substack{k',n'\\k,n}} \tilde{f}_{s'}(t,x,\eps',\theta')(1-\bar{f}_{s}(t,x,\eps,\theta)) \right.\\
& - \left. A_{s,s'}^{\substack{k,n\\k',n'}} \tilde{f}_{s}(t,x,\eps,\theta)(1- \tilde{f}_{s'}(t,x,\eps',\theta')) \right]\\
& = \sum_{s'} \sum_{k'=1}^{N_\eps} \sum_{n'=1}^{N_\theta} \int_{x_{i-1/2}}^{x_{i+1/2}} dx \,  \tilde{v}^m \left\lbrace A_{s',s}^{\substack{k',n'\\k,n}} \left[ a_{i,k',n'}^{s'}(t) + \frac{2(x-x_i)}{\Delta x} b_{i,k',n'}^{s'}(t)\right]\left[ 1- a_{i,k,n}^{s}(t) - \frac{2(x-x_i)}{\Delta x} b_{i,k,n}^{s}(t)\right] \right.\\
& - \left. A_{s,s'}^{\substack{k,n\\k',n'}} \left[ a_{i,k,n}^{s}(t) + \frac{2(x-x_i)}{\Delta x} b_{i,k,n}^{s}(t)\right]\left[ 1- a_{i,k',n'}^{s'}(t) - \frac{2(x-x_i)}{\Delta x} b_{i,k',n'}^{s'}(t)\right] \right\rbrace\\
& = \sum_{s'} \sum_{k'=1}^{N_\eps} \sum_{n'=1}^{N_\theta} \int_{x_{i-1/2}}^{x_{i+1/2}} dx \,  \tilde{v}^m \left\lbrace \left[A_{s',s}^{\substack{k',n'\\k,n}} a_{i,k',n'}^{s'}(t) \left( 1- a_{i,k,n}^{s}(t) \right) - A_{s,s'}^{\substack{k,n\\k',n'}} a_{i,k,n}^{s}(t) \left( 1- a_{i,k',n'}^{s'}(t) \right)\right] \right.\\
& - \frac{2(x-x_i)}{\Delta x} \left[ A_{s',s}^{\substack{k',n'\\k,n}} \left(  a_{i,k',n'}^{s'}(t)  b_{i,k,n}^{s}(t) - \left(1-a_{i,k,n}^{s}(t)\right)b_{i,k',n'}^{s'}(t) \right) \right.\\
&\left. - A_{s,s'}^{\substack{k,n\\k',n'}}  \left(  a_{i,k,n}^{s}(t)  b_{i,k',n'}^{s'}(t) - \left(1-a_{i,k',n'}^{s'}(t)\right)b_{i,k,n}^{s}(t) \right)\right]\\
& \left. - \frac{4(x-x_i)^2}{\Delta x^2} \left[ A_{s',s}^{\substack{k',n'\\k,n}} b_{i,k',n'}^{s'}(t)b_{i,k,n}^{s}(t)-A_{s,s'}^{\substack{k,n\\k',n'}} b_{i,k,n}^{s}(t)b_{i,k',n'}^{s'}(t) \right] \right\rbrace.
\end{align*}
\endgroup
Specifically, for $m=0$ we get
\begin{equation*}
\begin{aligned}
&\int_{C_{i,k,n}} \frac{\eps}{(\hbar v_F)^2} Q( \tilde{f}_s, \tilde{f}_{-s}) \, dx \, d\eps \, d\theta \\
&= \sum_{s'} \sum_{k'=1}^{N_\eps} \sum_{n'=1}^{N_\theta} \Delta x \left\lbrace \left[A_{s',s}^{\substack{k',n'\\k,n}} a_{i,k',n'}^{s'}(t) \left( 1- a_{i,k,n}^{s}(t) \right) - A_{s,s'}^{\substack{k,n\\k',n'}} a_{i,k,n}^{s}(t) \left( 1- a_{i,k',n'}^{s'}(t) \right)\right] \right.\\
& \left. - \frac{1}{3} \left[ A_{s',s}^{\substack{k',n'\\k,n}} b_{i,k',n'}^{s'}(t)b_{i,k,n}^{s}(t)-A_{s,s'}^{\substack{k,n\\k',n'}} b_{i,k,n}^{s}(t)b_{i,k',n'}^{s'}(t) \right] \right\rbrace,
\end{aligned}
\end{equation*}
and for $m=1$ we obtain
\begin{equation*}
\begin{aligned}
&\int_{C_{i,k,n}} \frac{2(x-x_i)}{\Delta x}\frac{\eps}{(\hbar v_F)^2} Q( \tilde{f}_s, \tilde{f}_{-s}) \, dx \, d\eps \, d\theta \\
&= - \frac{\Delta x}{3}\sum_{s'} \sum_{k'=1}^{N_\eps} \sum_{n'=1}^{N_\theta}  \left[ A_{s',s}^{\substack{k',n'\\k,n}} \left(  a_{i,k',n'}^{s'}(t)  b_{i,k,n}^{s}(t) - \left(1-a_{i,k,n}^{s}(t)\right)b_{i,k',n'}^{s'}(t) \right) \right.\\
&\left. - A_{s,s'}^{\substack{k,n\\k',n'}}  \left(  a_{i,k,n}^{s}(t)  b_{i,k',n'}^{s'}(t) - \left(1-a_{i,k',n'}^{s'}(t)\right)b_{i,k,n}^{s}(t) \right)\right].
\end{aligned}
\end{equation*}

We remark that a piece-wise linear DG discretization with respect to the energy and the angle leads to the computation of 16 families of coefficients with a structure similar to $A_{s',s}^{\substack{k',n'\\k,n}}$ for each $s$ and $s'$ \cite{NaCaRo}.

Considering all the terms discussed above, we obtain a set of $4 N_x N_\eps N_\theta$ ordinary differential equations. For the time discretization we adopt a third order TVD Runge-Kutta method \cite{ShuOsher}.

We remark that the electron distribution functions must be between 0 and 1 \cite{NaBoRo}. This cannot be automatically guaranteed  by adopting the DG discretization described above. To overcame this issue we adopt the maximum-principle-satisfying scheme introduced in \cite{ZhangShu}. It consists of a linear scaling around the cell average of the reconstruction. If
\begin{equation}
\tilde{f}_s(t,x,\eps,\theta) = a_{i,k,n}^s(t) + \frac{2(x-x_i)}{\Delta x} b_{i,k,n}^s(t), \qquad \forall(x,\eps,\theta)\in C_{i,k,n},
\end{equation}
is the reconstruction of $f_s$ in $C_{i,k,n}$ than we define
\begin{equation}
\tilde{\tilde{f}}_s(t,x,\eps,\theta) = a_{i,k,n}^s(t) + \vartheta\frac{2(x-x_i)}{\Delta x} b_{i,k,n}^s(t), \qquad \forall(x,\eps,\theta)\in C_{i,k,n},
\end{equation}
where $\vartheta$ is
\begin{equation}
\vartheta = \min\left\lbrace \frac{\left\vert 1-a_{i,k,n}^s(t) \right\vert}{\left\vert M-a_{i,k,n}^s(t) \right\vert}, \frac{\left\vert a_{i,k,n}^s(t) \right\vert}{\left\vert m-a_{i,k,n}^s(t) \right\vert}, 1 \right\rbrace
\end{equation}
and $M$, $m$ are the maximum and the minimum of $\tilde{f}_s(t,x,\eps,\theta)$ in $C_{i,k,n}$. Since the reconstruction is linear they are
\begin{equation}
\begin{aligned}
M & = \max\left\lbrace a_{i,k,n}^s(t)-b_{i,k,n}^s(t), a_{i,k,n}^s(t)+b_{i,k,n}^s(t) \right\rbrace,\\
m & = \min\left\lbrace a_{i,k,n}^s(t)-b_{i,k,n}^s(t), a_{i,k,n}^s(t)+b_{i,k,n}^s(t) \right\rbrace.
\end{aligned}
\end{equation}

To compute the macroscopic physical quantities we evaluate the cell averages with respect to $x$, which is equivalent to the piecewise constant reconstruction. For electrons we obtain
\begin{equation}
\begin{alignedat}{2}
n_i(t) & = n(t,x_i) \approx \frac{g_s g_v}{(2\pi)^2} \sum_{k=1}^{N_\eps}\sum_{n=1}^{N_\theta} a_{i,k,n}^+ \bar{N}_{k,n}, && \qquad \mbox{(electron density)}\\
j^{n,x}_i(t) & = j^{n,x}(t,x_i) \approx -e\frac{g_s g_v}{(2\pi)^2} \sum_{k=1}^{N_\eps}\sum_{n=1}^{N_\theta} a_{i,k,n}^+ \bar{R}_{k,n}, && \qquad \mbox{(electron current density)}\\
\mathcal{E}^n_i(t) & = \mathcal{E}^n(t,x_i) \approx \frac{g_s g_v}{(2\pi)^2} \sum_{k=1}^{N_\eps}\sum_{n=1}^{N_\theta} a_{i,k,n}^+ \bar{T}_{k,n}. && \qquad \mbox{(electron energy density)}
\end{alignedat}
\end{equation}
The coefficients $\bar{N}_{k,n}$, $\bar{R}_{k,n}$ and $\bar{T}_{k,n}$ are the integrals of the weight functions evaluated on the generic cell of the momentum space:
\begin{equation}
\begin{aligned}
\bar{N}_{k,n} & = \frac{\Delta\theta}{2(\hbar v_F)^2}\left(\eps_{k+\frac{1}{2}}^2-\eps_{k-\frac{1}{2}}^2\right),\\
\bar{R}_{k,n} & = \frac{v_F}{2(\hbar v_F)^2}\left(\eps_{k+\frac{1}{2}}^2-\eps_{k-\frac{1}{2}}^2\right)\left(\sin\theta_{n+\frac{1}{2}}-\sin\theta_{n-\frac{1}{2}}\right),\\
\bar{T}_{k,n} & = \frac{1}{3(\hbar v_F)^2}\left(\eps_{k+\frac{1}{2}}^3-\eps_{k-\frac{1}{2}}^3\right).
\end{aligned}
\end{equation}
Similarly, for holes we have
\begin{equation}
\begin{alignedat}{2}
p_i(t) & = p(t,x_i) \approx \frac{g_s g_v}{(2\pi)^2} \sum_{k=1}^{N_\eps}\sum_{n=1}^{N_\theta} (1-a_{i,k,n}^-) \bar{N}_{k,n}, && \qquad \mbox{(hole density)}\\
j^{p,x}_i(t) & = j^{p,x}(t,x_i) \approx e\frac{g_s g_v}{(2\pi)^2} \sum_{k=1}^{N_\eps}\sum_{n=1}^{N_\theta} (1-a_{i,k,n}^-) \bar{R}_{k,n}, && \qquad \mbox{(hole current density)}\\
\mathcal{E}^p_i(t) & = \mathcal{E}^p(t,x_i) \approx \frac{g_s g_v}{(2\pi)^2} \sum_{k=1}^{N_\eps}\sum_{n=1}^{N_\theta} (1-a_{i,k,n}^-) \bar{T}_{k,n}. && \qquad \mbox{(hole energy density)}
\end{alignedat}
\end{equation}

For plotting reason, the average quantities are also computed at $x=0$ and $x=L$ by using the boundary conditions.

To compute the right hand side of the Poisson equation at the edges of the mesh, i.e. $h_{i+\frac{1}{2},j+\frac{1}{2}}$, we first calculate the electron and hole density both at the left and right reconstruction of the distribution,
\begin{equation}
\begin{aligned}
n_{i+\frac{1}{2}}^- & = \frac{g_s g_v}{(2\pi)^2} \sum_{k=1}^{N_\eps}\sum_{n=1}^{N_\theta} (a_{i,k,n}^+ +b_{i,k,n}^+) \bar{N}_{k,n},\\
n_{i+\frac{1}{2}}^+ & = \frac{g_s g_v}{(2\pi)^2} \sum_{k=1}^{N_\eps}\sum_{n=1}^{N_\theta} (a_{i+1,k,n}^+ -b_{i+1,k,n}^+) \bar{N}_{k,n},\\
p_{i+\frac{1}{2}}^- & = \frac{g_s g_v}{(2\pi)^2} \sum_{k=1}^{N_\eps}\sum_{n=1}^{N_\theta} [1-(a_{i,k,n}^- +b_{i,k,n}^-)] \bar{N}_{k,n},\\
p_{i+\frac{1}{2}}^+ & = \frac{g_s g_v}{(2\pi)^2} \sum_{k=1}^{N_\eps}\sum_{n=1}^{N_\theta} [1-(a_{i+1,k,n}^- -b_{i+1,k,n}^-)] \bar{N}_{k,n},
\end{aligned}
\end{equation}
and then average the quantities at the two sides of each edge 
\begin{equation}
\begin{aligned}
n_{i+\frac{1}{2}} & = \frac{1}{2}\left( n_{i+\frac{1}{2}}^- + n_{i+\frac{1}{2}}^+ \right),\\
p_{i+\frac{1}{2}} & = \frac{1}{2}\left( p_{i+\frac{1}{2}}^- + p_{i+\frac{1}{2}}^+ \right).
\end{aligned} 
\end{equation}

The overall the numerical discretization is summarized in Algorithm \ref{alg:dg_boltzmann}.

\begin{algorithm}[H]
\caption{Overall solver of the BTE}
\label{alg:dg_boltzmann}
\begin{algorithmic}[1]

\State \textbf{Initialization:} 
Construct the DG approximation of the initial distribution function, obtaining $a^s_{i,k,n}(0)$ and $b^s_{i,k,n}(0)$. 

Initialize the electrostatic potential and compute electric field $E_x$.

Precompute collision coefficents $A_{s',s}^{\substack{k',n'\\k,n}}$.

\For{$r = 0,1,\dots,N_t-1$}

    \State \textbf{Compute the transport in physical space.} 

    \State \textbf{Compute the transport in momentum space.} 

    \State \textbf{Compute the collision term.} 

    \State \textbf{Time integration.} 
    Combine the previous substeps using a Runge--Kutta method to obtain $a^s_{i,k,n}(t_{r+1})$ and $b^s_{i,k,n}(t_{r+1})$.

    \State \textbf{Positivity-preserving limiter.} 
    Enforce non-negativity of the numerical solution by applying a suitable limiter.

    \State \textbf{Poisson solve.} 
    Compute the electrostatic potential from the charge density and update the electric field.

\EndFor

\end{algorithmic}
\end{algorithm}

\section{Simulation results}
\label{SEC:Sim_res}
In this section we use the numerical approach presented above to simulate the cases described  in Sec. \ref{SEC:Phys_sett}, first investigating the accuracy of the 
 numerical method adopted.
\subsection{Test case 1: suspended monolayer graphene}
    The study of the convergence of the numerical scheme has been addressed comparing the simulation results of the macroscopic quantities at stationary regime on different meshes with respect to the variables $x$, $\eps$ and $\theta$. The analysis is performed in the physical situation of the test case of Fig. \ref{FIG:Cont_gr}. The device length is 100 nm and the Fermi level in the boundary conditions is set to 0.25 eV. As obtained in  \cite{MajNaRo} with this value of the Fermi level the hole current is negligible compared to the electron contribution. Since the graphene layer is suspended, we do not consider the electron-substrate collision term $Q^{(el-sub)}$. We adopt the same physical parameters as those used in Section \ref{SEC:Sim_res}, with the electric field set to 1 V/$\mu$m. The time at which we compare the results is 1 ps, when the stationary regime is already reached.

First we discuss the convergence of the scheme with respect to $x$ analyzing  electron density, mean velocity, and mean energy. We consider a fixed discretization of $N_\eps=80$ and $N_\theta=32$ in the energy-angle space. Let $U_i^{\Delta x}$ with $i=1,2,\ldots,N_x$ the cell average of the generic macroscopic quantity adopting a uniform discretization with respect to $x$ of step size $\Delta x$. Halving the step size, the corresponding quantity is $U_i^{\Delta x/2}$ with $i=1,2,\ldots,2N_x$. Following \cite{Toro}, to compare the solutions on the two meshes, we adopt a similar approach employed for finite volumes schemes. We set
$$
\tilde{U}_i^{\Delta x/2} = \frac{1}{2}\left[ U_{2i-1}^{\Delta x/2}+U_{2i}^{\Delta x/2} \right], \qquad i=1,2,\ldots,N_x,
$$ 
which is the cell average computed with the finer solution on the coarser mesh. We define the error as
$$
\text{Err\,}_U^p(\Delta x) = \norm{U^{\Delta x}-\tilde{U}^{\Delta x/2}}_p,
$$
where $\norm{\cdot}_p$ indicates the $L^p\left([0,L]\right)$ norm. Consequently, the convergence rate $\alpha_U^p$ is
$$
\alpha_U^p = \log_2\left( \frac{\text{Err\,}_U^p(\Delta x)}{\text{Err\,}_U^p(\Delta x/2)} \right).
$$
We summarize the obtained results in the Table \ref{TAB:Num_st_x}. Even if a full second order is not reached, the results indicate that in the adopted norms a convergence rate definitely higher than one is obtained.

Regarding the convergence of the scheme with respect to $\eps$ and $\theta$ we consider a fixed discretization with respect to $x$ with $N_x=40$ cells. Let $U_i^{\Delta \eps}$ and $U_i^{\Delta \theta}$ with $i=1,2,\ldots,N_x$ the cell average of the generic macroscopic quantity adopting a uniform discretization with respect to $\eps$ and $\theta$ of step size $\Delta \eps$ and $\Delta \theta$, respectively. The errors are defined as
$$
\text{Err\,}_U^p(\Delta \eps) = \norm{U^{\Delta \eps}-U^{\Delta \eps/2}}_p \qquad \text{and} \qquad \text{Err\,}_U^p(\Delta \theta) = \norm{U^{\Delta \theta}-U^{\Delta \theta/2}}_p.
$$
Moreover, when the convergence study is done with respect to $\eps$ we consider a fixed discretization with respect to $\theta$ of $N_\theta=32$ cells; when the convergence study is done with respect to $\theta$ we consider a fixed discretization with respect to $\eps$ of $N_\eps=40$ cells. The convergence results with respect to $\eps$ and $\theta$ are reported in Tables \ref{TAB:Num_st_eps} and \ref{TAB:Num_st_th}, respectively. The results are similar to those for spatial mesh but a bit worse, especially for the $\theta$-mesh. In any case an order greater than one is obtained. The lesser accuracy in $\eps$ and $\theta$ presumably is to ascribe to the  adoption of elements which are linear in $x$ but constant with respect to other variables. Our choice represents  a good compromise between accuracy and computational complexity.   

\begin{table}[!ht]
\centering
\setlength{\tabcolsep}{8pt} 
\renewcommand{\arraystretch}{1.2} 
\begin{tabular}{|c|c|c|c|c|c|c|c|}
\hline 
\rule[-1ex]{0pt}{2.5ex} $U$ & $N_x$ & $\text{Err\,}_U^1$ & $\alpha_U^1$ & $\text{Err\,}_U^2$ & $\alpha_U^2$ & $\text{Err\,}_U^\infty$ & $\alpha_U^\infty$ \\ 
\hline 
\rule[-1ex]{0pt}{2.5ex} \multirow{4}{*}{$n$} & 40 & 4.2677e$-$01 &  & 3.0449e$+$00 &  & 3.1555e$+$01 &  \\ 
\rule[-1ex]{0pt}{2.5ex} & 80 & 6.3712e$-$02 & 2.7435 & 5.2056e$-$01 & 2.5483 & 9.2600e$+$00 & 1.7688 \\ 
\rule[-1ex]{0pt}{2.5ex} & 160 & 1.7837e$-$02 & 1.8367 & 1.4056e$-$01 & 1.8889 & 2.9350e$+$00 & 1.6577 \\
\cline{2-8}
\rule[-1ex]{0pt}{2.5ex} & slope & & 2.2901 &  & 2.2186 & & 1.7132 \\
\hline
\rule[-1ex]{0pt}{2.5ex} \multirow{4}{*}{$V^{x,n}$} & 40 & 3.3900e$-$06 &  & 2.4872e$-$05 &  & 2.6115e$-$04 &  \\ 
\rule[-1ex]{0pt}{2.5ex} & 80 & 5.2712e$-$07 & 2.6851 & 4.1572e$-$06 & 2.5808 & 7.4100e$-$05 & 1.8173 \\ 
\rule[-1ex]{0pt}{2.5ex} & 160 & 1.4269e$-$07 & 1.8853 & 1.1403e$-$06 & 1.8662 & 2.3800e$-$05 & 1.6385 \\
\cline{2-8}
\rule[-1ex]{0pt}{2.5ex} & slope & & 2.2852 & & 2.2235 & & 1.7279 \\
\hline
\rule[-1ex]{0pt}{2.5ex} \multirow{4}{*}{$E^n$} & 40 & 7.6050e$-$07 &  & 4.9671e$-$06 &  & 5.6150e$-$05 &  \\ 
\rule[-1ex]{0pt}{2.5ex} & 80 & 1.1925e$-$07 & 2.6730 & 7.0638e$-$07 & 2.8139 & 9.3000e$-$06 & 2.5940 \\ 
\rule[-1ex]{0pt}{2.5ex} & 160 & 3.0312e$-$08 & 1.9760 & 1.8218e$-$07 & 1.9551 & 3.1500e$-$06 & 1.5619 \\
\cline{2-8}
\rule[-1ex]{0pt}{2.5ex} & slope & & 2.3245 & & 2.3845 & & 2.0779 \\
\hline
\end{tabular}
\caption{Convergence rate with respect to $x$ keeping fixed $N_\eps=80$ and $N_\theta=32$.}\label{TAB:Num_st_x}
\end{table}

\begin{table}[!ht]
\centering
\setlength{\tabcolsep}{8pt} 
\renewcommand{\arraystretch}{1.2} 
\begin{tabular}{|c|c|c|c|c|c|c|c|}
\hline 
\rule[-1ex]{0pt}{2.5ex} $U$ & $N_x$ & $\text{Err\,}_U^1$ & $\alpha_U^1$ & $\text{Err\,}_U^2$ & $\alpha_U^2$ & $\text{Err\,}_U^\infty$ & $\alpha_U^\infty$ \\ 
\hline 
\rule[-1ex]{0pt}{2.5ex} \multirow{4}{*}{$n$} & 40 & 3.1633e$+$01 &  & 1.0448e$+$02 &  & 4.6315e$+$02 &  \\ 
\rule[-1ex]{0pt}{2.5ex} & 80 & 8.2431e$+$00 & 1.9402 & 2.7416$+$01 & 1.9301 & 1.2676e$+$02 & 1.8694 \\ 
\rule[-1ex]{0pt}{2.5ex} & 160 & 1.9766e$+$00 & 2.0602 & 6.5795e$+$00 & 2.0590 & 3.1370e$+$01 & 2.0146 \\
\cline{2-8}
\rule[-1ex]{0pt}{2.5ex} & slope & & 2.0002 &  & 1.9946 & & 1.9420 \\
\hline
\rule[-1ex]{0pt}{2.5ex} \multirow{4}{*}{$V^{x,n}$} & 40 & 5.3858e$-$04 &  & 1.7839e$-$03 &  & 8.2872e$-$03 &  \\ 
\rule[-1ex]{0pt}{2.5ex} & 80 & 1.5670e$-$04 & 1.7811 & 5.1889e$-$04 & 1.7815 & 2.4323e$-$03 & 1.7686 \\ 
\rule[-1ex]{0pt}{2.5ex} & 160 & 3.8789e$-$05 & 2.0143 & 1.2843e$-$06 & 2.0145 & 6.0800e$-$04 & 2.0146 \\
\cline{2-8}
\rule[-1ex]{0pt}{2.5ex} & slope & & 1.8977 & & 1.8980 & & 1.8844 \\
\hline
\rule[-1ex]{0pt}{2.5ex} \multirow{4}{*}{$E^n$} & 40 & 3.9385e$-$04 &  & 1.2562e$-$03 &  & 4.4673e$-$03 &  \\ 
\rule[-1ex]{0pt}{2.5ex} & 80 & 1.2016e$-$04 & 1.7127 & 3.8373e$-$04 & 1.7109 & 1.3848e$-$03 & 1.6897 \\ 
\rule[-1ex]{0pt}{2.5ex} & 160 & 3.2771e$-$05 & 1.8745 & 1.0465e$-$04 & 1.8745 & 3.7820e$-$04 & 1.8725 \\
\cline{2-8}
\rule[-1ex]{0pt}{2.5ex} & slope & & 1.7936 & & 1.7927 & & 1.7811 \\
\hline
\end{tabular}
\caption{Convergence rate with respect to $\eps$ keeping fixed $N_x=40$ and $N_\theta=32$.}\label{TAB:Num_st_eps}
\end{table}

\begin{table}[!ht]
\centering
\setlength{\tabcolsep}{8pt} 
\renewcommand{\arraystretch}{1.2} 
\begin{tabular}{|c|c|c|c|c|c|c|c|}
\hline 
\rule[-1ex]{0pt}{2.5ex} $U$ & $N_x$ & $\text{Err\,}_U^1$ & $\alpha_U^1$ & $\text{Err\,}_U^2$ & $\alpha_U^2$ & $\text{Err\,}_U^\infty$ & $\alpha_U^\infty$ \\ 
\hline 
\rule[-1ex]{0pt}{2.5ex} \multirow{4}{*}{$n$} & 32 & 3.1998e$+$01 &  & 1.0712e$+$02 &  & 6.4936e$+$02 &  \\ 
\rule[-1ex]{0pt}{2.5ex} & 64 & 1.1621e$+$00 & 1.4612 & 3.8530$+$01 & 1.4752 & 2.5946e$+$02 & 1.3235 \\ 
\rule[-1ex]{0pt}{2.5ex} & 128 & 4.3586e$+$00 & 1.4148 & 1.4288e$+$01 & 1.4312 & 1.0544e$+$02 & 1.2991 \\
\cline{2-8}
\rule[-1ex]{0pt}{2.5ex} & slope & & 1.4380 &  & 1.4532 & & 1.3113 \\
\hline
\rule[-1ex]{0pt}{2.5ex} \multirow{4}{*}{$V^{x,n}$} & 32 & 2.0632e$-$04 &  & 7.6983e$-$04 &  & 5.5621e$-$03 &  \\ 
\rule[-1ex]{0pt}{2.5ex} & 64 & 7.3795e$-$05 & 1.4833 & 2.7428e$-$04 & 1.4889 & 2.2769e$-$03 & 1.2886 \\ 
\rule[-1ex]{0pt}{2.5ex} & 128 & 2.6938e$-$05 & 1.4539 & 9.8183e$-$05 & 1.4821 & 8.8980e$-$04 & 1.3555 \\
\cline{2-8}
\rule[-1ex]{0pt}{2.5ex} & slope & & 1.4686 & & 1.4855 & & 1.3220 \\
\hline
\rule[-1ex]{0pt}{2.5ex} \multirow{4}{*}{$E^n$} & 32 & 5.9165e$-$05 &  & 1.9778e$-$04 &  & 1.1366e$-$03 &  \\ 
\rule[-1ex]{0pt}{2.5ex} & 64 & 2.0734e$-$05 & 1.5127 & 6.8979e$-$05 & 1.5196 & 4.8840e$-$04 & 1.2186 \\ 
\rule[-1ex]{0pt}{2.5ex} & 128 & 7.7378e$-$06 & 1.4220 & 2.5337e$-$05 & 1.4449 & 1.7300e$-$04 & 1.4973 \\
\cline{2-8}
\rule[-1ex]{0pt}{2.5ex} & slope & & 1.4674 & & 1.4822 & & 1.3579 \\
\hline
\end{tabular}
\caption{Convergence rate with respect to $\theta$ keeping fixed $N_x=40$ and $N_\eps=40$.}\label{TAB:Num_st_th}
\end{table}

In the sequel a numerical mesh of $N_x=80$, $N_\eps=100$, and $N_\theta=32$ cells is adopted. We consider two values of the external electric field, namely 1 V/$\mu$m and 2 V/$\mu$m. The stationary regime is reached in about 1 ps also for the electric field of 2 V/$\mu$m. 

\begin{figure}[!ht]
\centering
\includegraphics[width=0.45\textwidth]{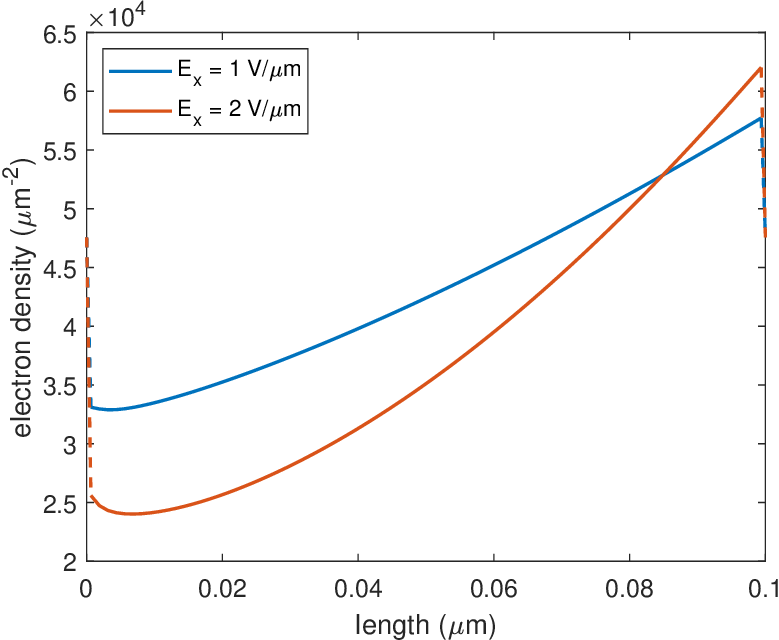}\qquad
\includegraphics[width=0.45\textwidth]{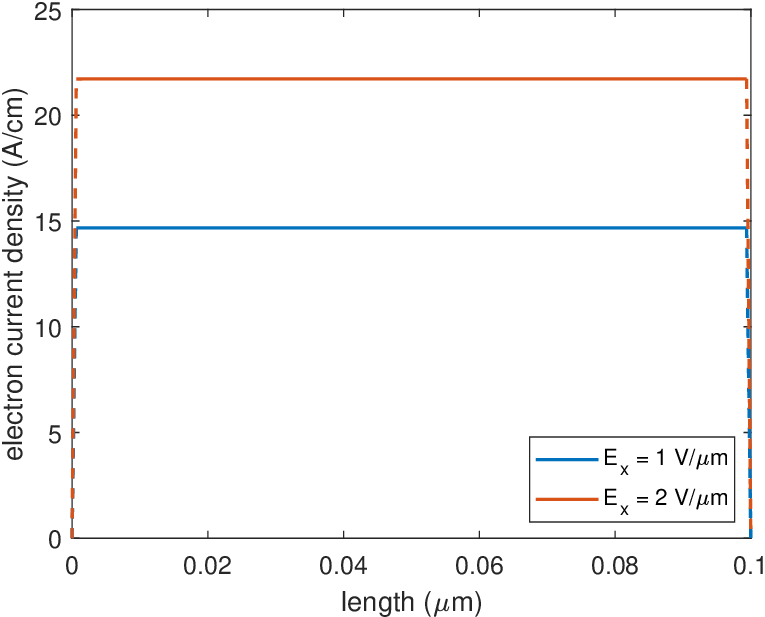}\\\medskip
\includegraphics[width=0.45\textwidth]{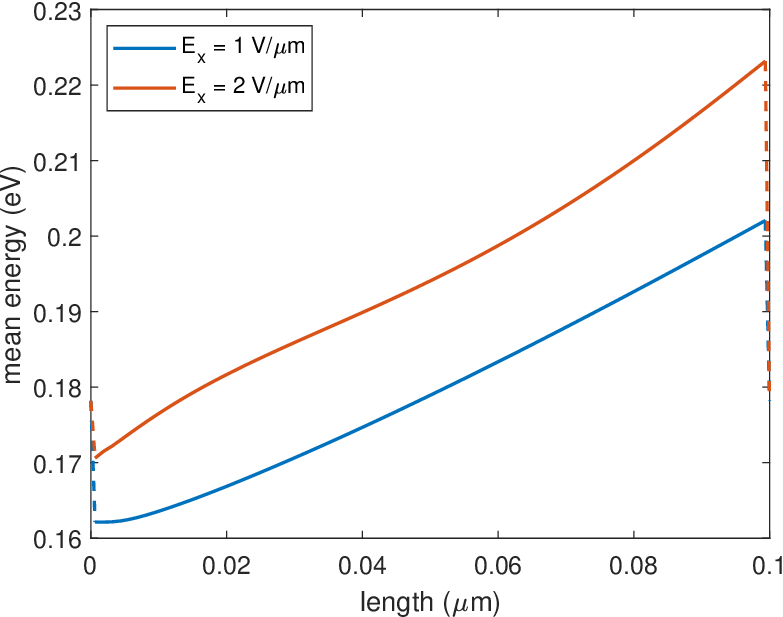}
\caption{Stationary electron density (top-left), current density (top-right) and mean energy (bottom) of the device schematized in Fig. \ref{FIG:Cont_gr} for applied fields of 1 V/$\mu$m and 2 V/$\mu$m.}\label{FIG:QOI_hom_cont}
\end{figure}

In Fig. \ref{FIG:QOI_hom_cont} we show the stationary electron density, electron current density, and mean energy. At the boundaries a jump in such quantities, represented by the dashed lines, along the metal-semiconductor junction is observed. This can be ascribed to the different states available in the contact, which is considered as a reservoir, and the graphene; in fact, we have an abrupt passage from a 3D material, the metal, and a 2D material, the graphene.  The electron density presents a depletion area close to the source contacts and an accumulation region close to the drain contact. In steady state, the electron current density is constant along the device length. The mean energy exhibits a  qualitative behavior similar to the electron density. 

The boundary layers are not interpreted as ill-posedness, but as kinetic contact layers generated by the inflow reservoir boundary conditions imposed at the metal–graphene interfaces. In the model, the contacts inject carriers with a Fermi–Dirac distribution fixed by the metal work function, while the interior distribution is determined by transport, electric acceleration, and scattering. Thus the solution must rapidly connect two different kinetic states: the reservoir state at the contact and the self-consistent graphene transport state in the channel.

\begin{figure}[!ht]
\centering
\includegraphics[width=0.32\textwidth]{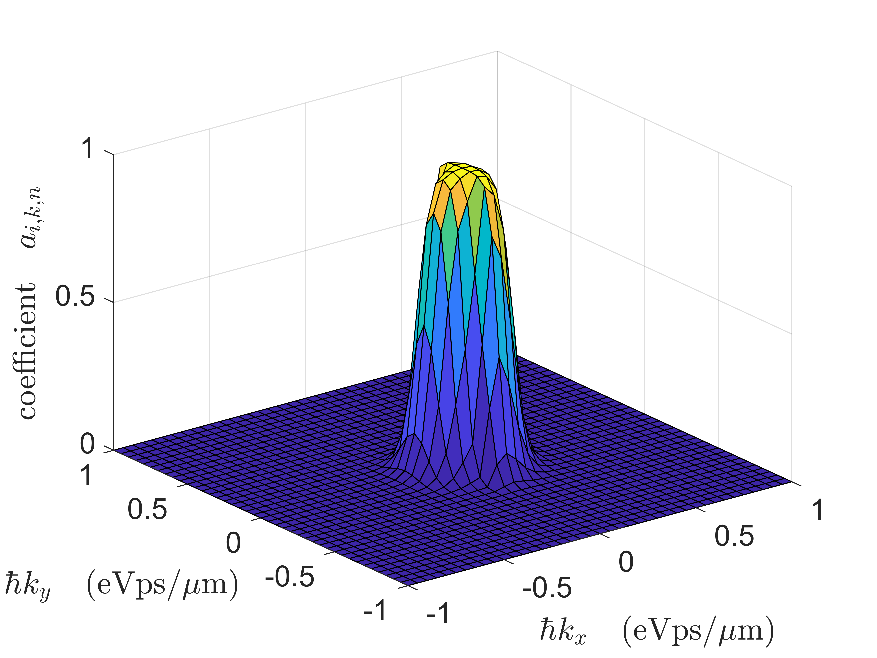}\,
\includegraphics[width=0.32\textwidth]{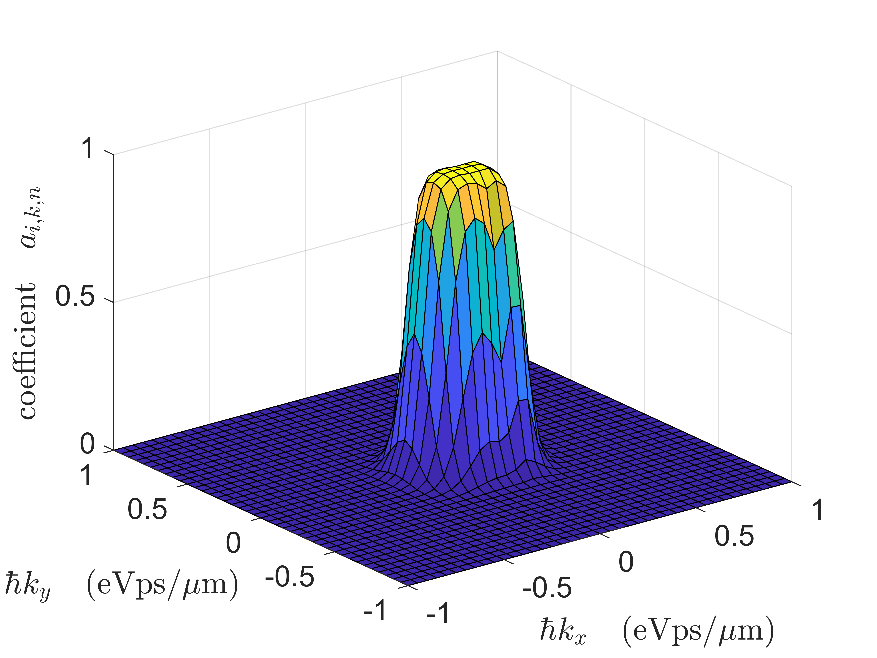}\,
\includegraphics[width=0.32\textwidth]{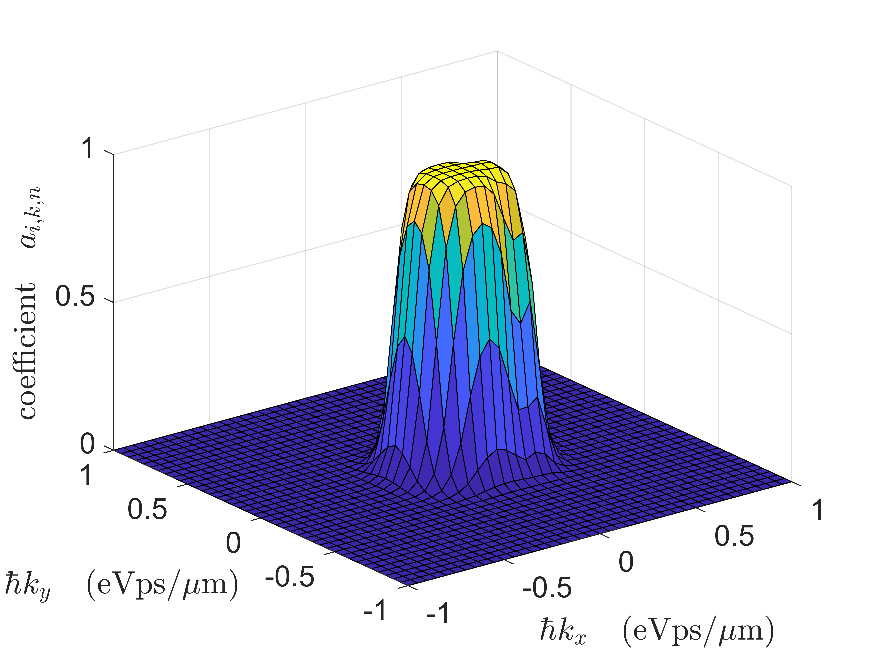}\\\medskip
\includegraphics[width=0.32\textwidth]{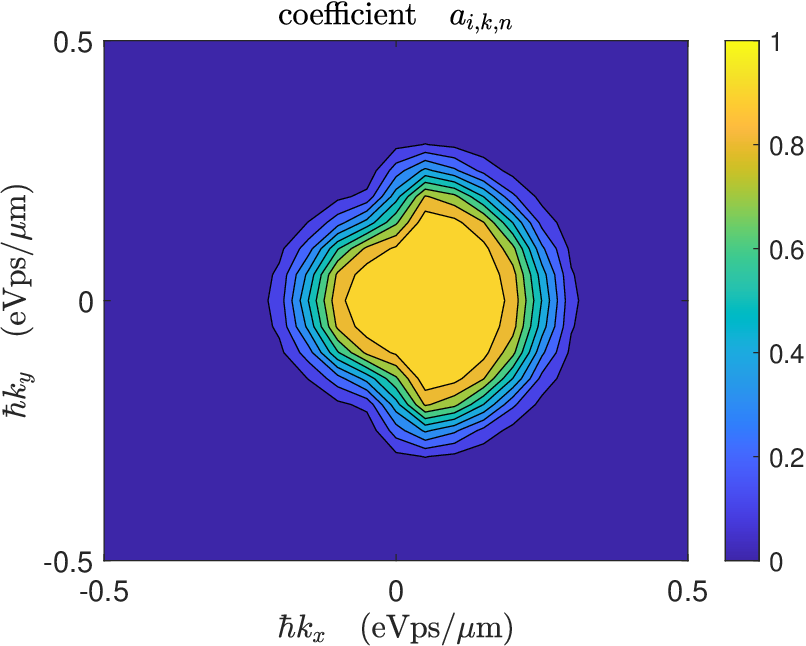}\,
\includegraphics[width=0.32\textwidth]{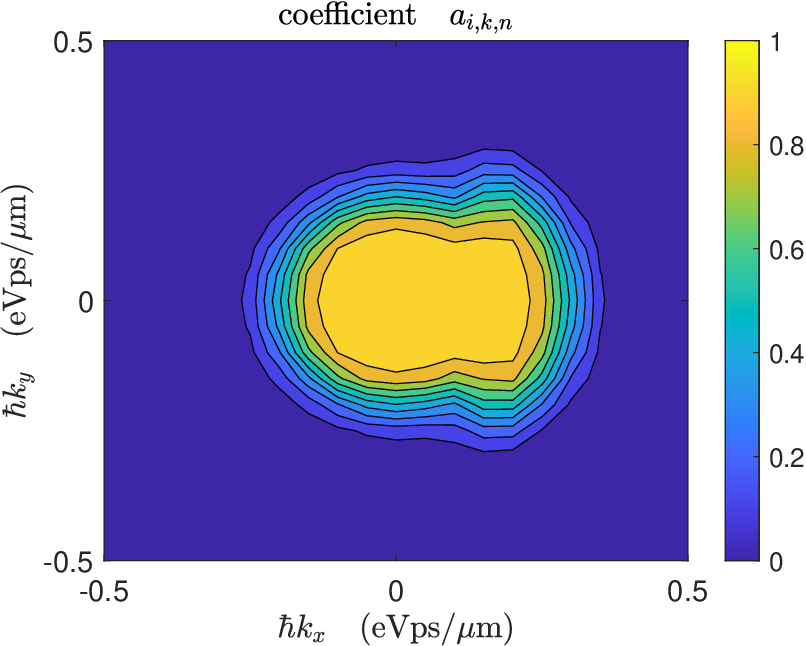}\,
\includegraphics[width=0.32\textwidth]{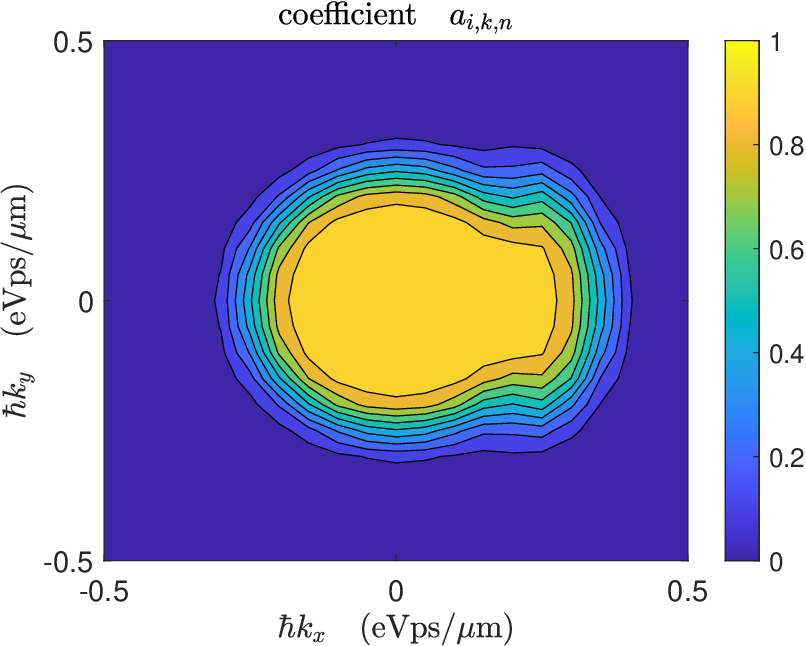}\\\medskip
\includegraphics[width=0.32\textwidth]{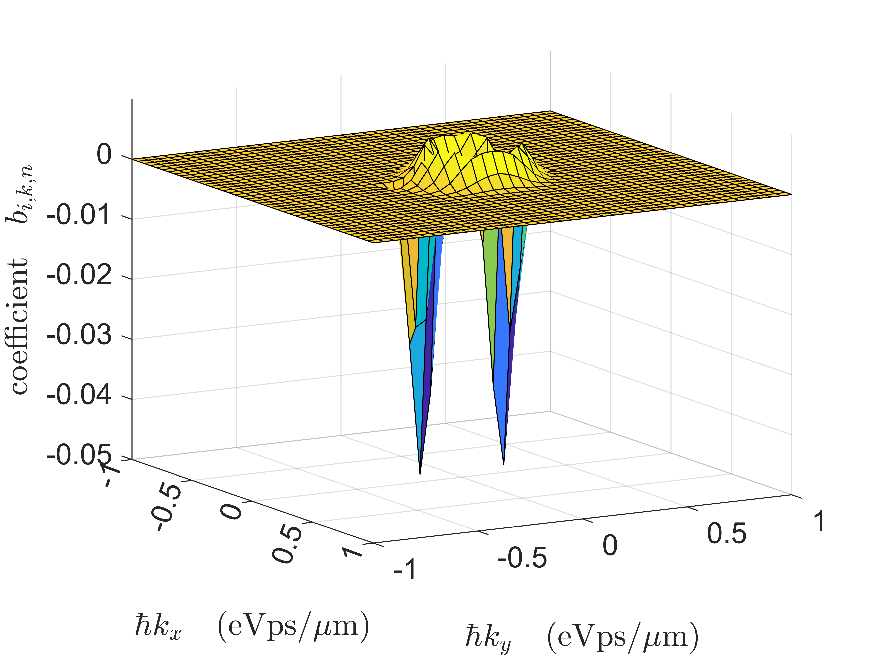}\,
\includegraphics[width=0.32\textwidth]{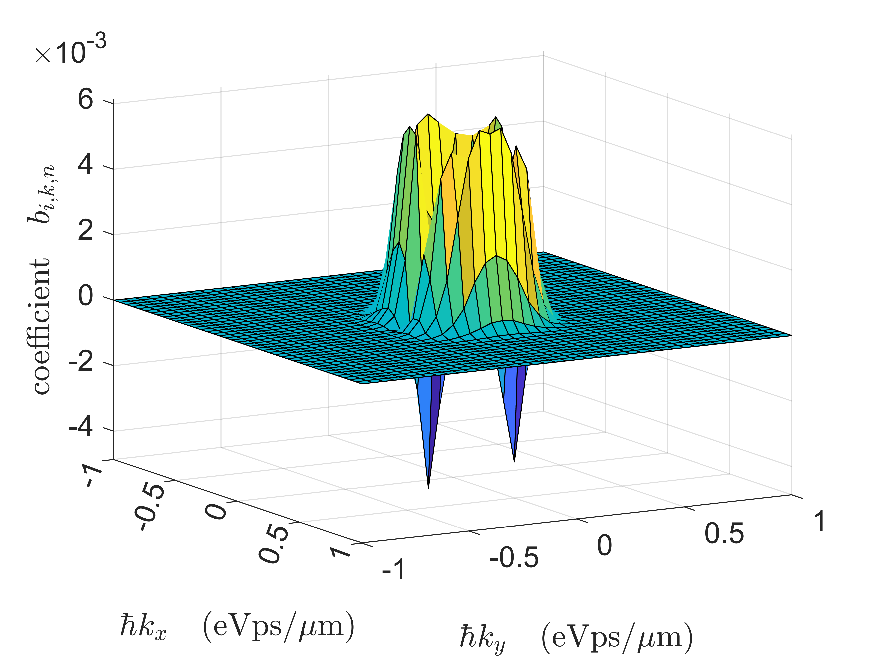}\,
\includegraphics[width=0.32\textwidth]{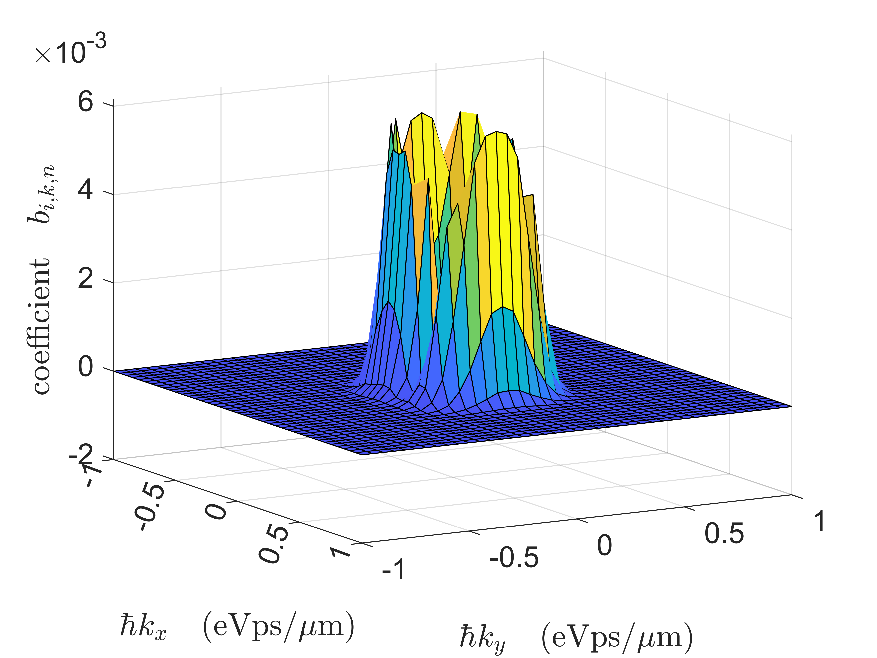}
\caption{Stationary $a_{i,k,n}$ coefficient of the approximated distribution function \eqref{EQ:FF_approx} at $x=0$ (top-left), $x=L/2$ (top-center) and $x=L$ (top-right) of the ideal device schematized in Fig. \ref{FIG:Cont_gr} in the conduction band. Contour plot of the $a_{i,k,n}$ coefficient at the same positions (central line plots). Stationary $b_{i,k,n}$ coefficient of the approximated distribution function at $x=0$ (bottom-left), $x=L/2$ (bottom-center) and $x=L$ (bottom-right). Applied field of 1 V/$\mu$m.}
\label{FIG:FF_hom_cont_1}
\end{figure}

\begin{figure}[!ht]
\centering
\includegraphics[width=0.32\textwidth]{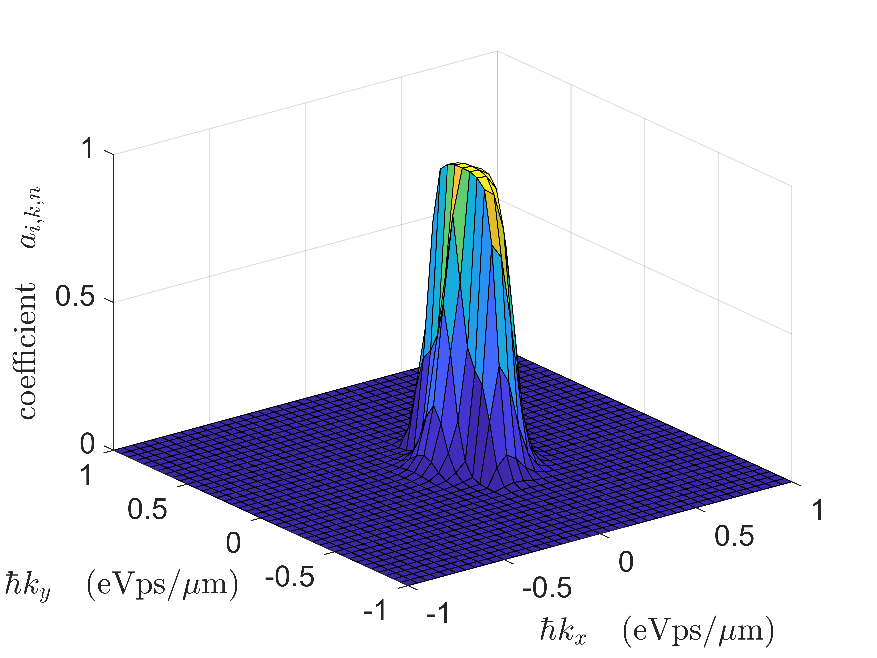}\,
\includegraphics[width=0.32\textwidth]{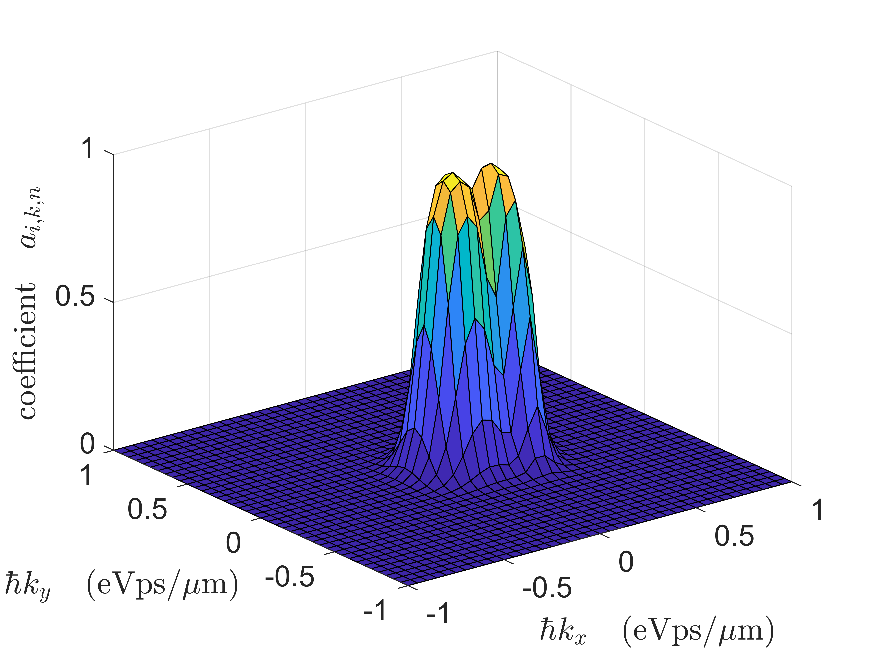}\,
\includegraphics[width=0.32\textwidth]{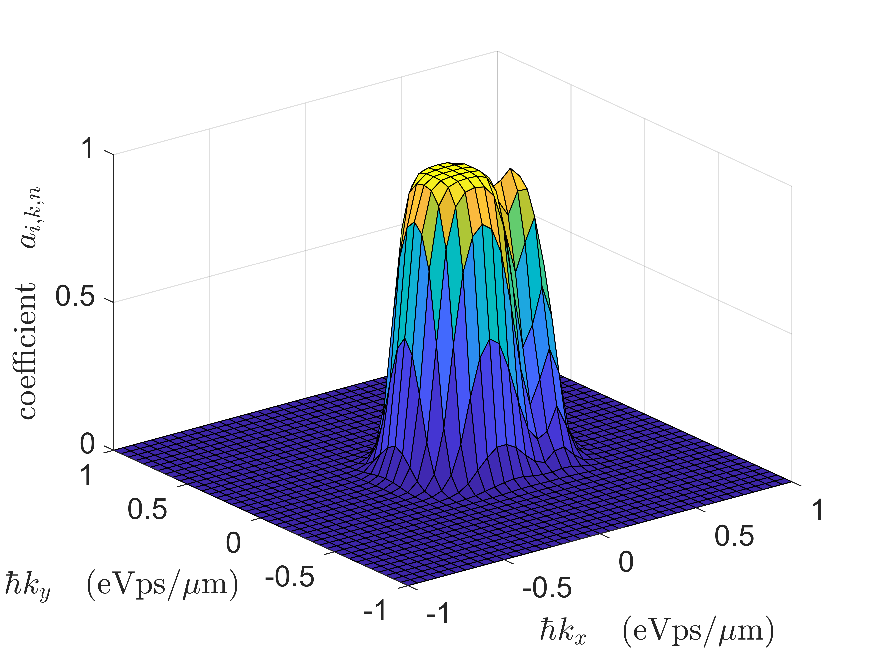}\\\medskip
\includegraphics[width=0.32\textwidth]{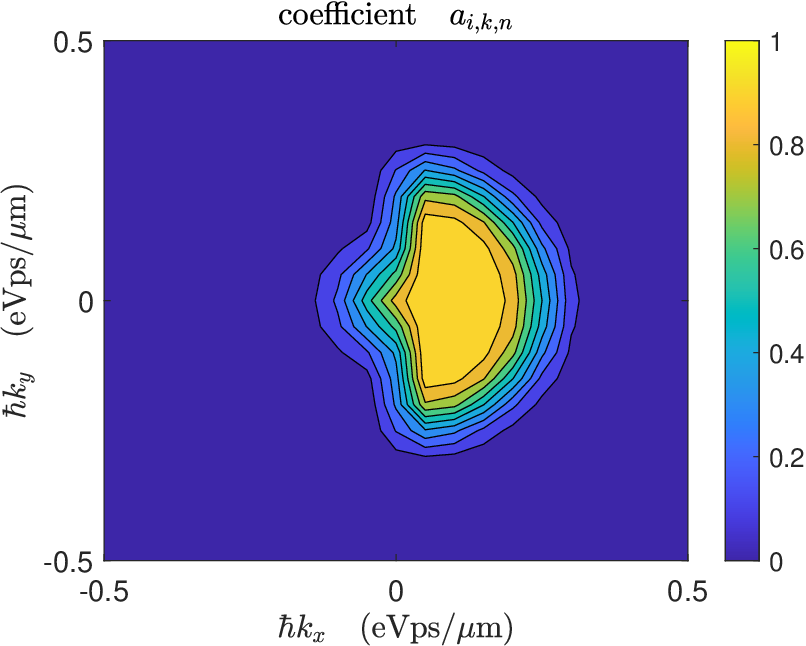}\,
\includegraphics[width=0.32\textwidth]{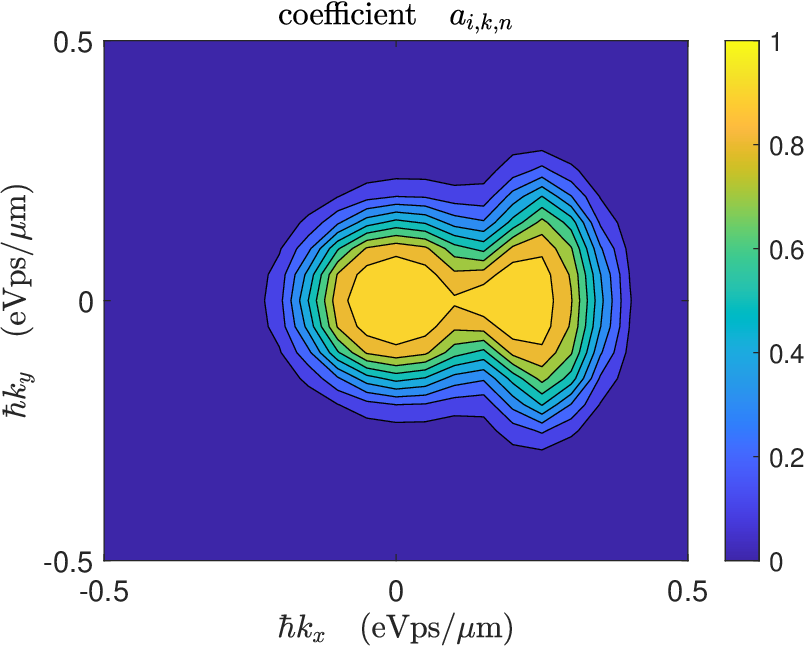}\,
\includegraphics[width=0.32\textwidth]{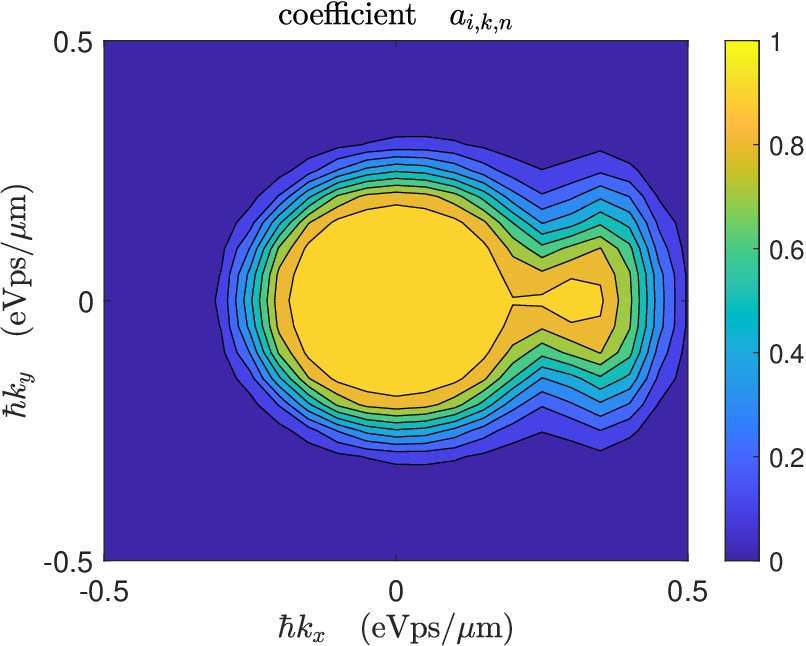}\\\medskip
\includegraphics[width=0.32\textwidth]{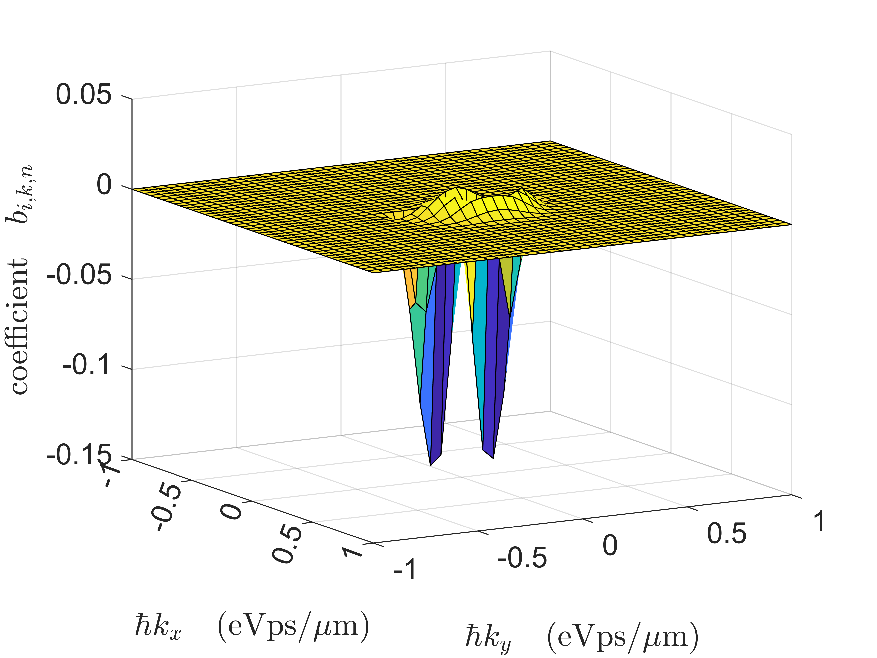}\,
\includegraphics[width=0.32\textwidth]{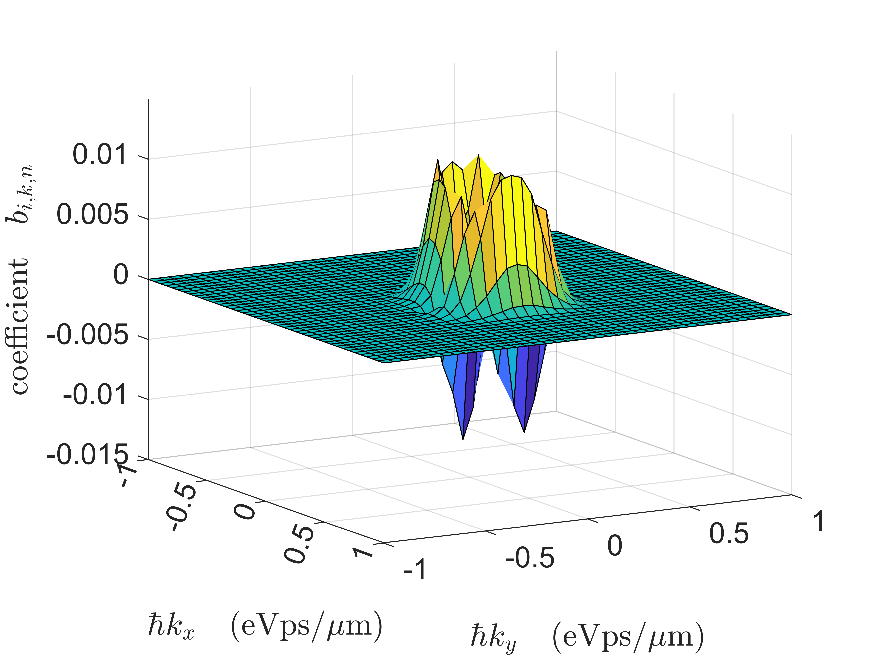}\,
\includegraphics[width=0.32\textwidth]{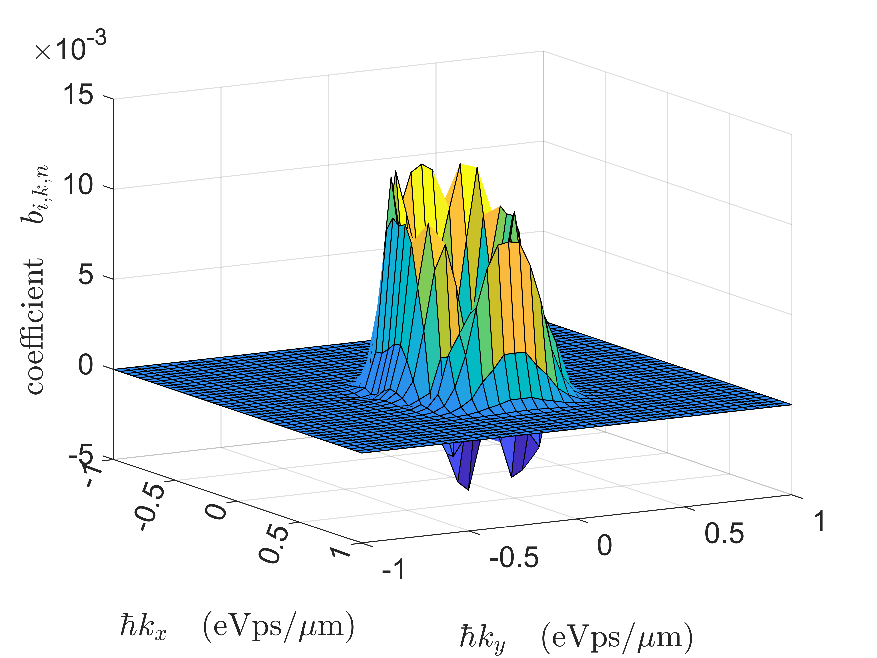}
\caption{Stationary $a_{i,k,n}$ coefficient of the approximated distribution function \eqref{EQ:FF_approx} at $x=0$ (top-left), $x=L/2$ (top-center) and $x=L$ (top-right) of the ideal device schematized in Fig. \ref{FIG:Cont_gr} in the conduction band. Contour plot of the $a_{i,k,n}$ coefficient at the same positions (central line plots). Stationary $b_{i,k,n}$ coefficient of the approximated distribution function at $x=0$ (bottom-left), $x=L/2$ (bottom-center) and $x=L$ (bottom-right). Applied field of 2 V/$\mu$m.}
\label{FIG:FF_hom_cont_2}.
\end{figure}

In Figs. \ref{FIG:FF_hom_cont_1}, \ref{FIG:FF_hom_cont_2} we show the stationary $a_{i,k,n}$ and $b_{i,k,n}$ coefficients of the approximated distribution function \eqref{EQ:FF_approx} in the conduction band. We observe that a second spike appears in the plot of the distribution function when the electric field is 2 V/$\mu$m while is missing for 1 V/$\mu$m. It is clear that such an effect is due to the strength of the electric field which accentuates the anisotropy of the distribution function.


\begin{figure}[!ht]
\centering
\includegraphics[width=0.45\textwidth]{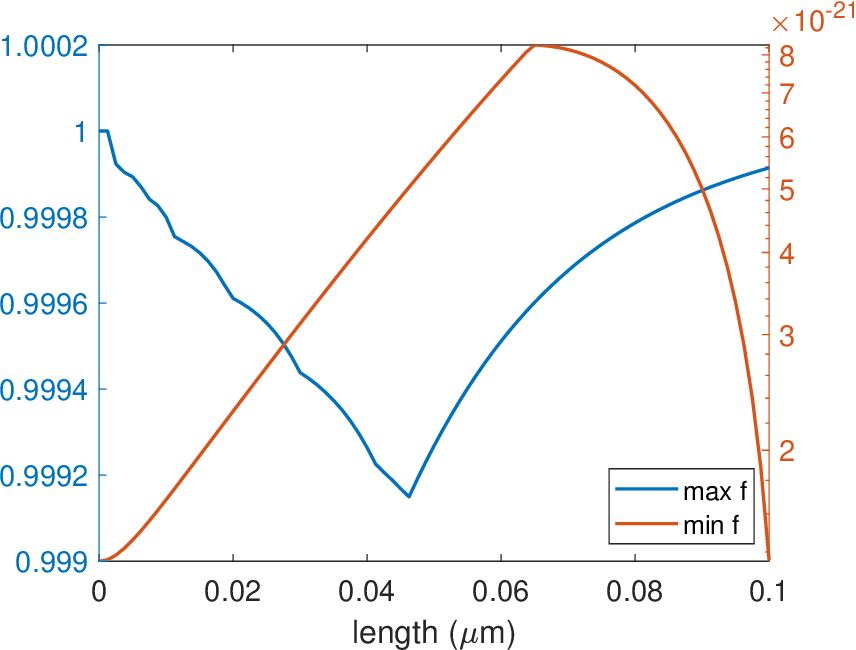}
\includegraphics[width=0.45\textwidth]{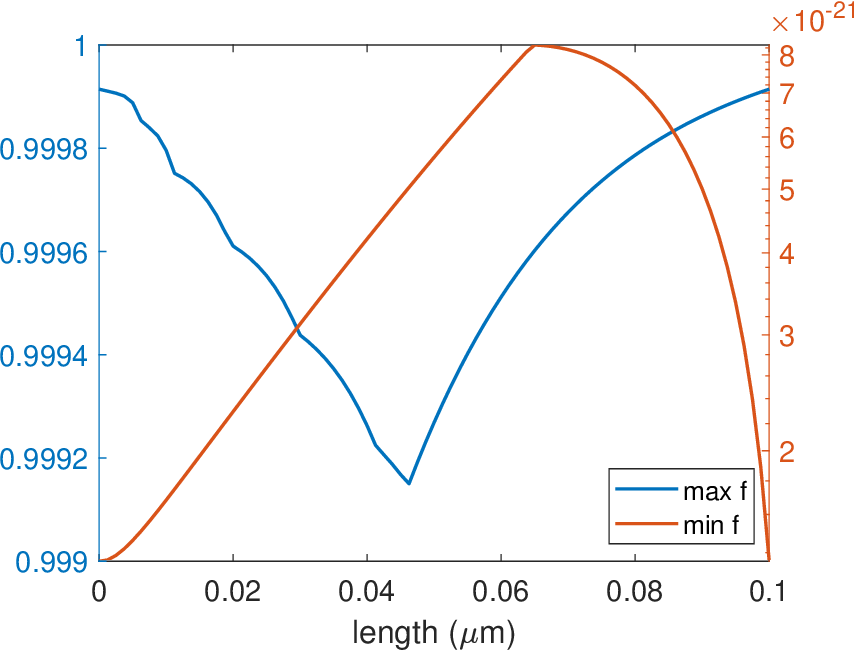}
\caption{Cellwise maxima and minima of the left and right reconstructed solutions.}\label{FIG:min_max_rec}
\end{figure}

Figure \ref{FIG:min_max_rec} presents a direct visualization of the maximum and the minimum of left and right reconstructed numerical solutions for each cell. The figure clearly shows that the reconstructed distributions remain within the physically admissible range, thereby validating the effectiveness of the maximum-principle-preserving limiter. In this case, the electric field is 1 V/$\mu$m.

\subsection{Test case 2: GFET}
The second test case is the simulation of the charge transport in the GFET described in Sec. \ref{SEC:Phys_sett} and schematized in Fig. \ref{FIG:GFET}. For the simulations, we consider $L = 100$ nm, $H=21$ nm, $V_G = 0.4$ V, and $V_b=0.1$ V. We adopt a numerical mesh of $N_x=80$, $N_\eps=100$, and $N_\theta=32$ cells. For the Poisson equation we discretize the domain in 81$\times$23 points. The stationary regime is reached in about 0.5 ps.

\begin{figure}[!ht]
\centering
\includegraphics[width=0.45\textwidth]{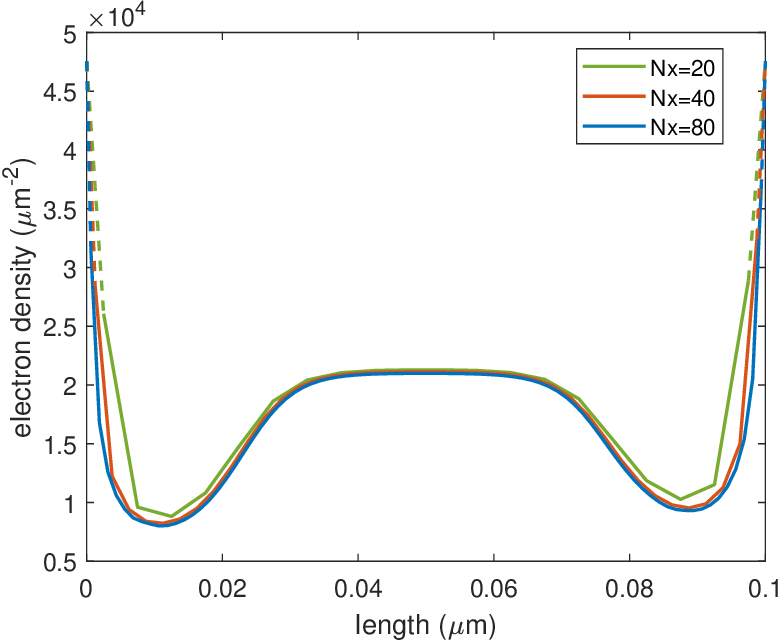}\qquad
\includegraphics[width=0.45\textwidth]{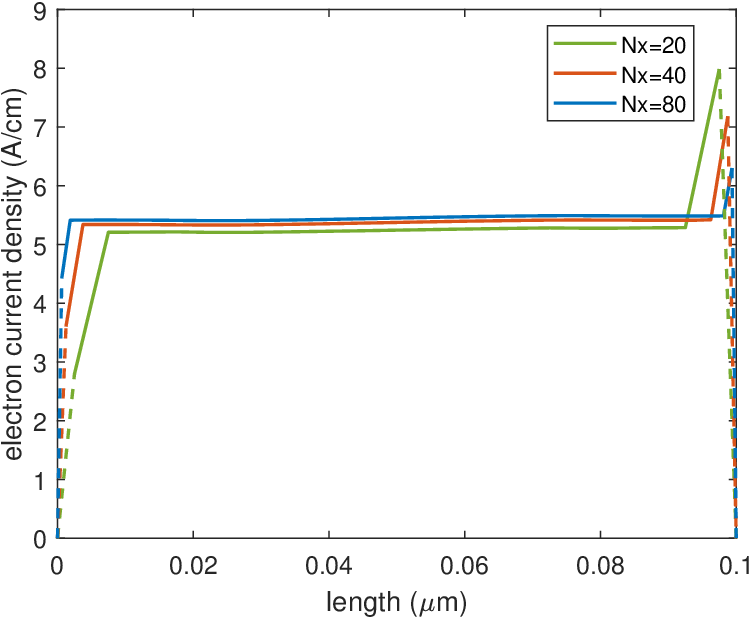}\\\medskip
\includegraphics[width=0.45\textwidth]{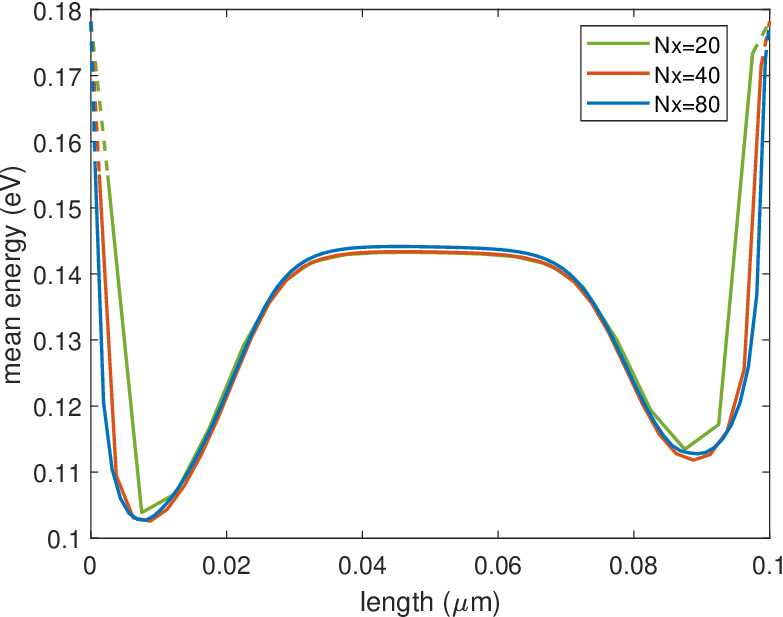}
\caption{Stationary electron density (top-left), current density (top-right) and mean energy (bottom) of the simulated GFET of Fig. \ref{FIG:GFET}.}\label{FIG:QOI_GFET}
\end{figure}

In Fig. \ref{FIG:QOI_GFET} we show the stationary electron density, the current density and the mean energy versus position of the simulated GFET. The different lines correspond to simulations performed with varying numbers of cells in the spatial mesh. Observe that the density presents a boundary layer close to the interface metal-graphene which is better resolved as the mesh is refined. The presence of such behavior can be explained with the same considerations of the test case 1.  The current is constant in the interior of the domain. The linear shape close to the contacts tends to diminish  as the mesh is refined. So, we can ascribe such an effect to 
the numerical approximation. 

\begin{figure}[!ht]
\centering
\includegraphics[width=0.32\textwidth]{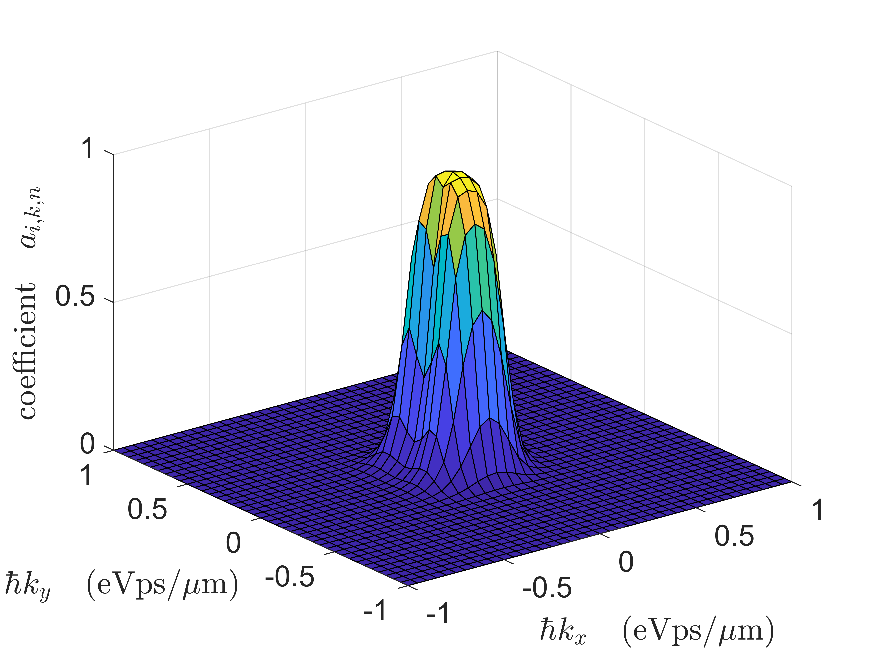}\,
\includegraphics[width=0.32\textwidth]{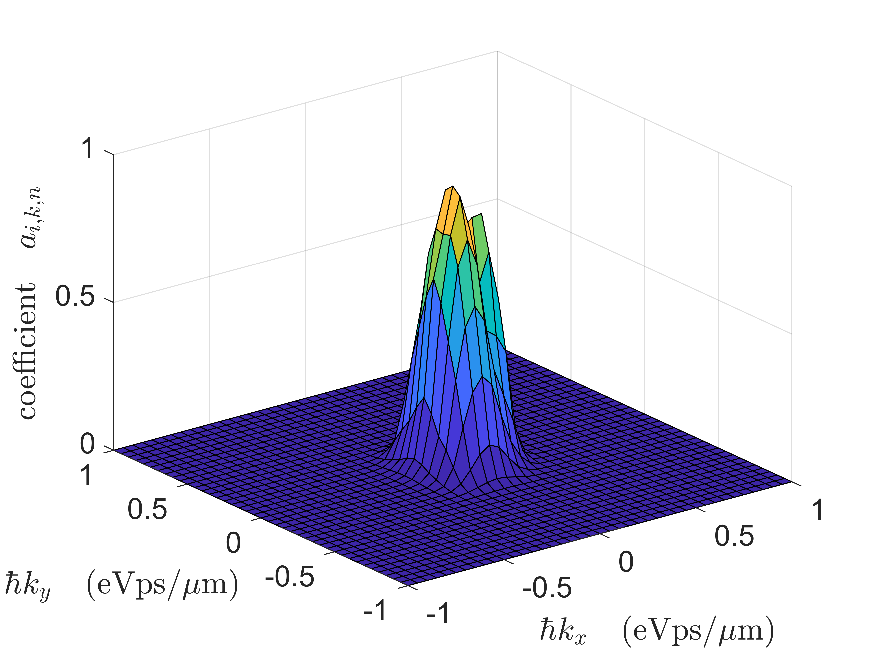}\,
\includegraphics[width=0.32\textwidth]{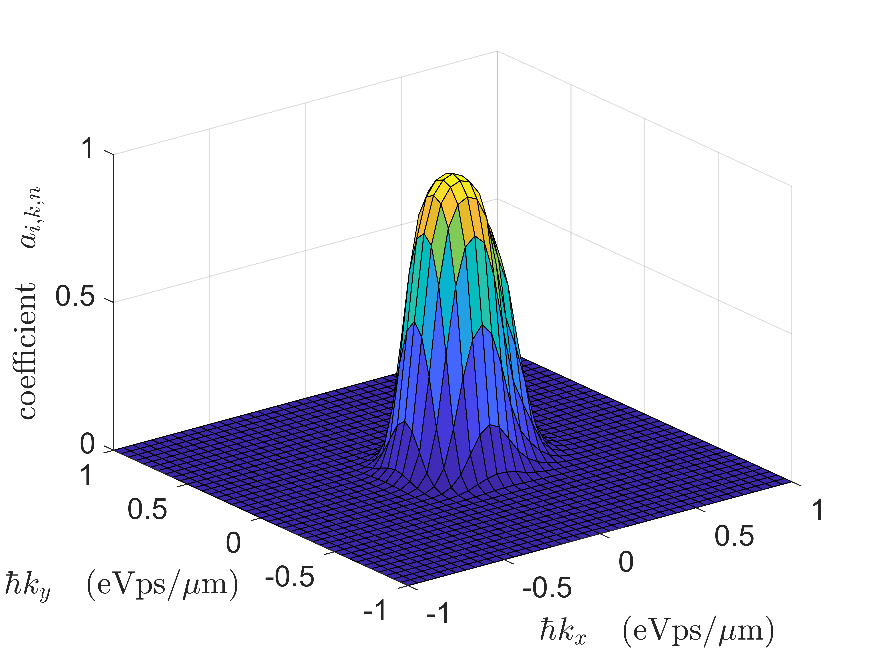}\\\medskip
\includegraphics[width=0.32\textwidth]{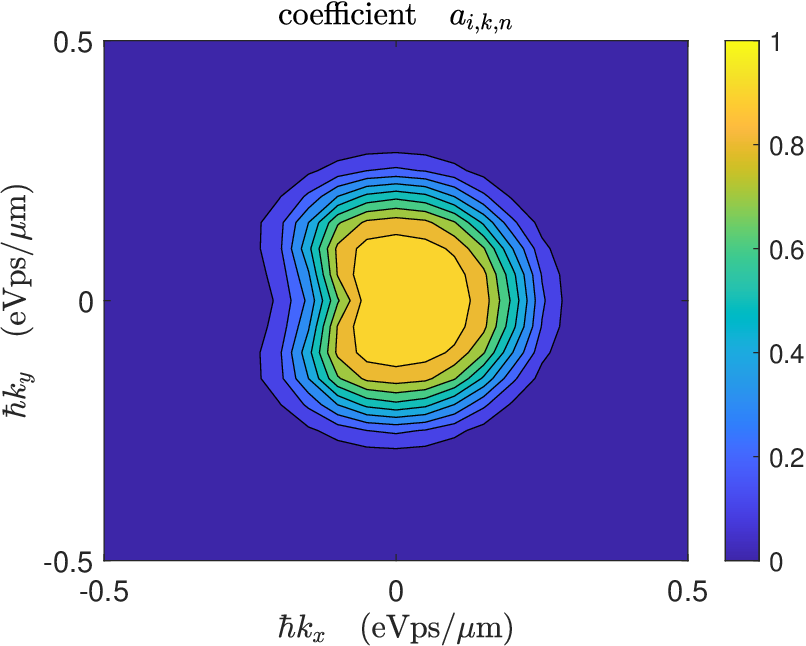}\,
\includegraphics[width=0.32\textwidth]{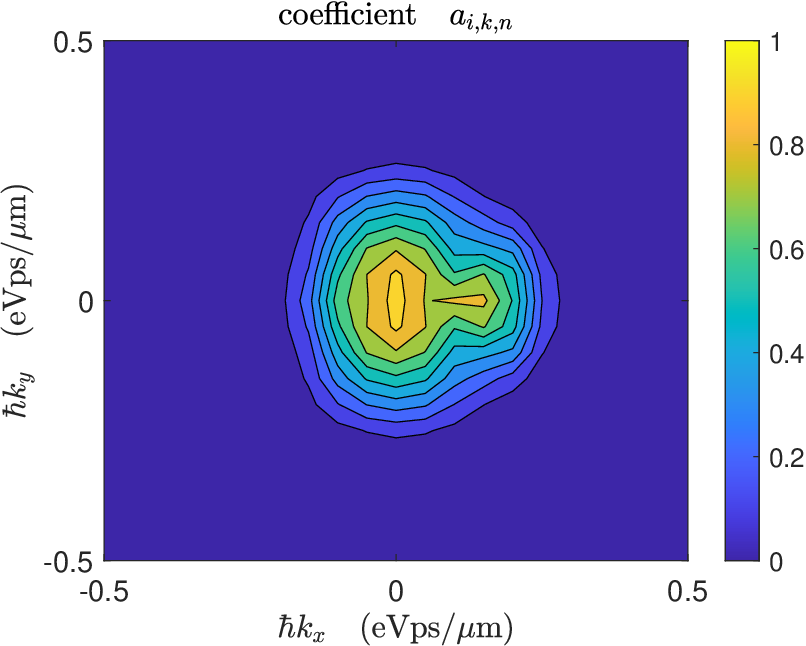}\,
\includegraphics[width=0.32\textwidth]{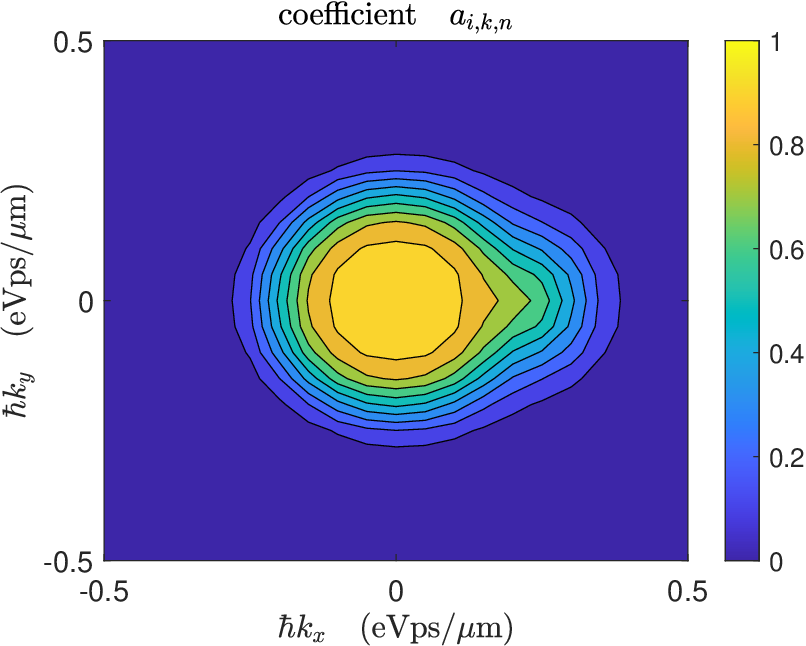}\\\medskip
\includegraphics[width=0.32\textwidth]{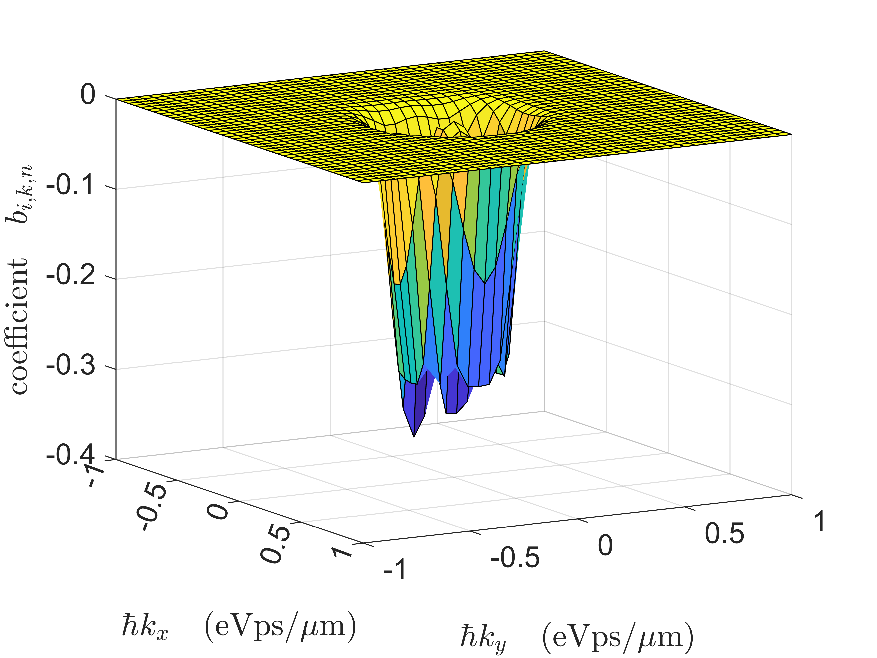}\,
\includegraphics[width=0.32\textwidth]{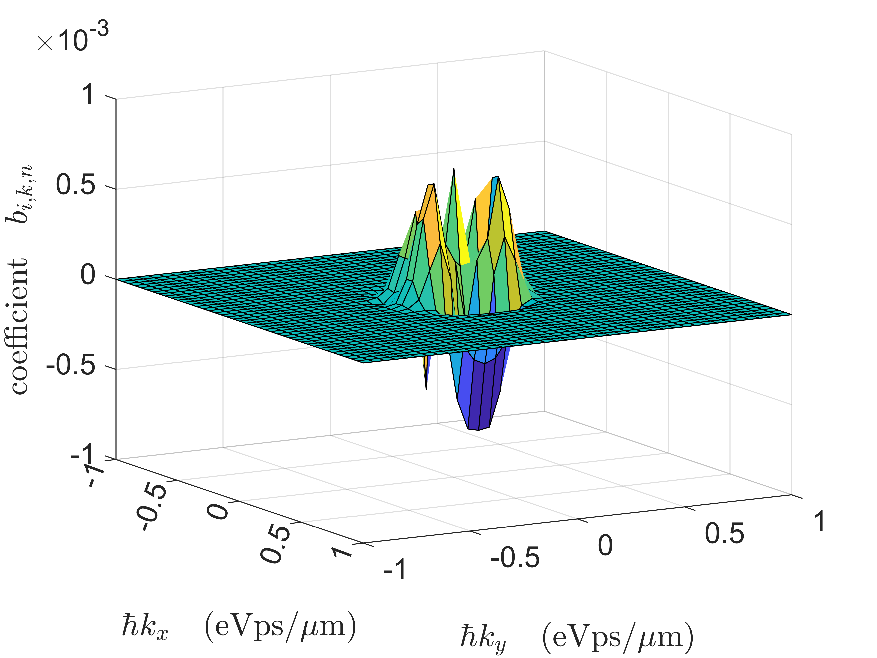}\,
\includegraphics[width=0.32\textwidth]{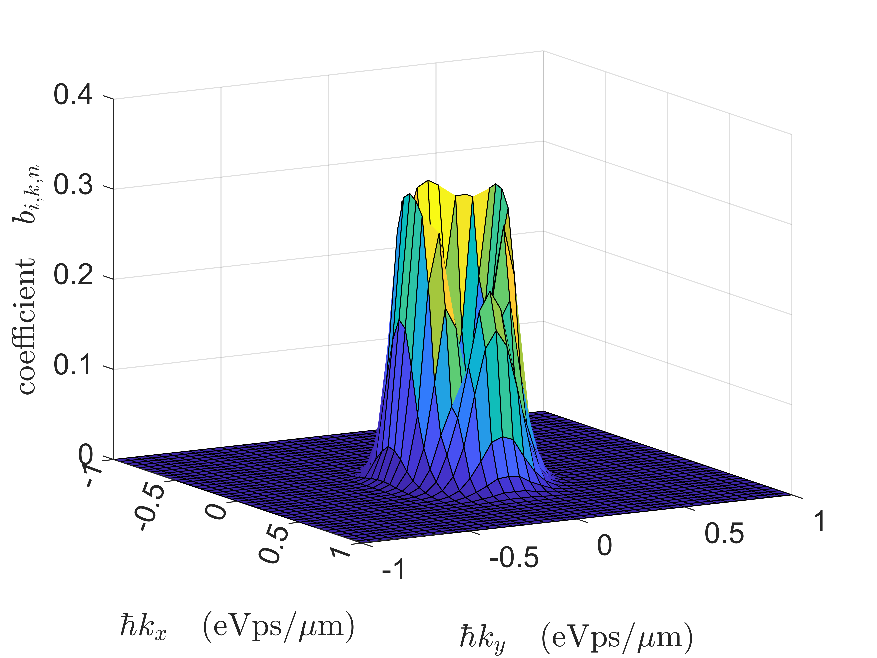}
\caption{Stationary $a_{i,k,n}$ coefficient of the approximated distribution function \eqref{EQ:FF_approx} at $x=0$ (top-left), $x=L/2$ (top-center) and $x=L$ (top-right) of the simulated GFET of Fig. \ref{FIG:GFET}. Contour plot of the $a_{i,k,n}$ coefficient at the same positions (central line plots). Stationary $b_{i,k,n}$ coefficient of the approximated distribution function at $x=0$ (bottom-left), $x=L/2$ (bottom-center) and $x=L$ (bottom-right). Applied field of 1 V/ $\mu$.}\label{FIG:FF_GFET}
\end{figure}

\begin{figure}[!ht]
\centering
\includegraphics[width=0.51\textwidth]{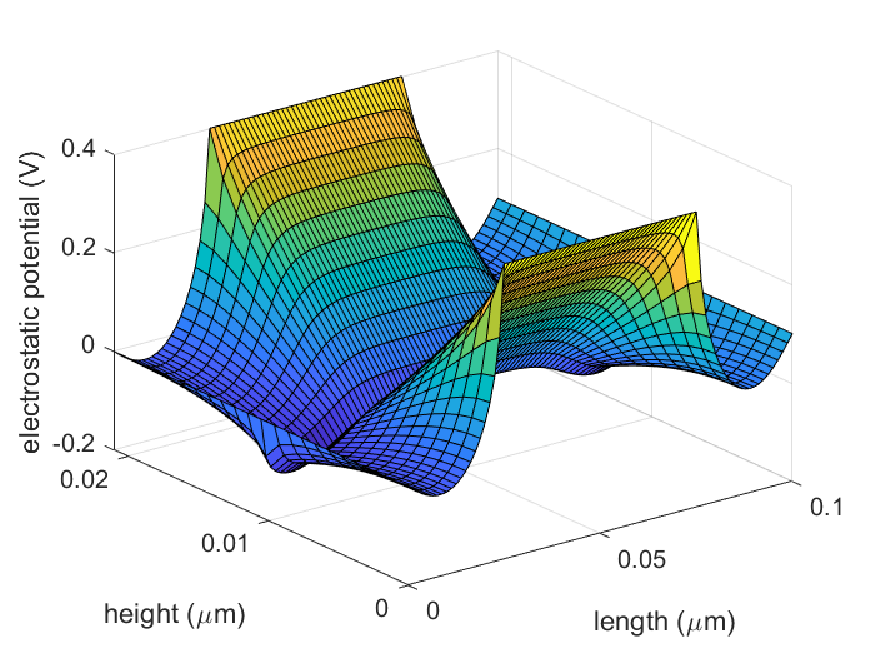}\qquad
\includegraphics[width=0.42\textwidth]{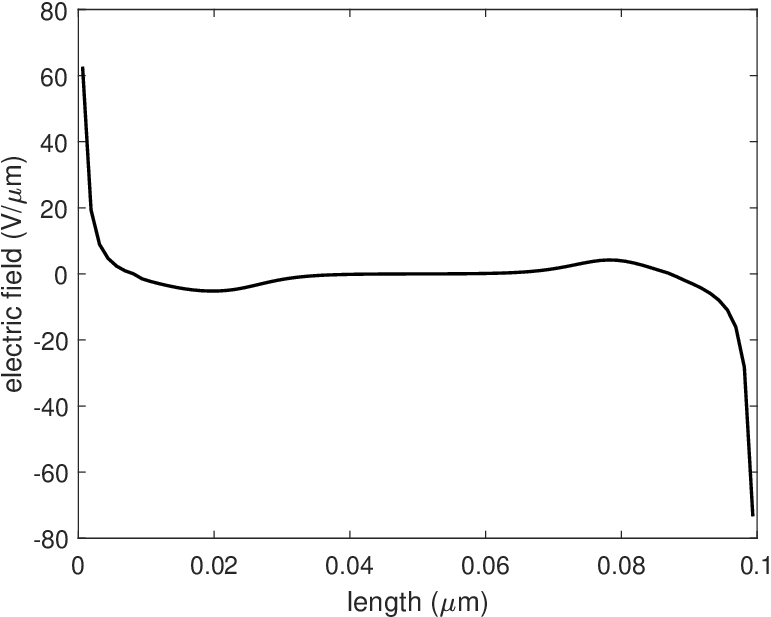}
\caption{Stationary electrostatic potential (left) and stationary electric field along the graphene layer (right) of the simulated GFET of Fig. \ref{FIG:GFET}.}\label{FIG:pot_GFET}
\end{figure}

Finally, in Fig. \ref{FIG:FF_GFET} we present the stationary $a_{i,k,n}$ and $b_{i,k,n}$ coefficients of the approximated distribution function \eqref{EQ:FF_approx}, referred to the simulated GFET while in Fig. \ref{FIG:pot_GFET} we show the stationary electrostatic potential, obtained solving the Poisson equation \eqref{EQ:Poisson} on the 2D device section, and the stationary electric field computed along the graphene layer, i.e. along $y=y_{gr}$. As expected, the higher anisotropy of the distribution function is in the middle of the graphene layer.

\section{Conclusions and acknowledgments}
In this work, we have obtained numerical simulations of a graphene field-effect transistors described directly by the semiclassical Boltzmann transport equation, employing a discontinuous Galerkin discretization with linear elements in space and constant approximation in the wave-vector. To guarantee that the distribution function satisfies the correct physical bounds, the maximum-principle-preserving scheme proposed in \cite{ZhangShu} has been adopted.

The method demonstrates robustness, with a good degree of accuracy and  well suited for capturing the complex charge transport dynamics inherent to graphene-based devices. The combination of the DG method with linear spatial elements ensures computational efficiency while preserving the essential physical features of the system, establishing the framework as a valuable tool for both theoretical investigations and device-level simulations of graphene transistors. 

The main novelty is the direct use of the semiclassical Boltzmann equation retaining the full complexity of the collision operator which includes all the electron-phonon scatterings along with the Pauli exclusion principle. We stress that in a Direct Simulation Monte Carlo the Pauli exclusion principle gives raise to controversial questions \cite{RoMajCo,BoCoDeRo}. The direct simulation with the DG method gives an elegant way to overcome such an issue.  

The kinetic approach reveals a new feature of the solution: the presence of a boundary layer at the contacts. The drift-diffusion equations, which are the model usually employed for the simulation of GFETs, do not catch such behaviour because the boundary conditions fix the density at the contacts \cite{NaRo_IEEE}.  The relevant role of the contact/graphene interface is instead crucial in boundary conditions for the Boltzmann equations. The formulation of a physically  more accurate  kinetic model for the contact/graphene interface
is currently under investigation by the authors and will be the subject of a forthcoming article.

\bigskip

The authors acknowledge the support from INdAM (GNFM) and from MUR progetto PRIN ``Transport phonema in low dimensional structures: models, simulations and theoretical aspects CUP E53D23005900006''.  V.R. acknowledges
also the support from the project PIACERI LINEA 1 \emph{TraPlas}, University of Catania.

\FloatBarrier
\end{document}